\documentclass[%
aip,
amsmath,amssymb,
reprint, onecolumn 
]{revtex4-1}

\usepackage{graphicx}
\usepackage{dcolumn}
\usepackage{bm}

\usepackage[utf8]{inputenc}
\usepackage[T1]{fontenc}
\usepackage{mathptmx}
\usepackage{etoolbox}
\usepackage{amsmath}
\usepackage{amssymb}
\makeatletter
\def\@email#1#2{%
	\endgroup
	\patchcmd{\titleblock@produce}
	{\frontmatter@RRAPformat}
	{\frontmatter@RRAPformat{\produce@RRAP{*#1\href{mailto:#2}{#2}}}\frontmatter@RRAPformat}
	{}{}
}%
\makeatother
\begin{document}
	
	\preprint{AIP/123-QED}
	
	\title{Two new super-integrable hierarchies and a generalized super-AKNS hierarchy}
	
	\author{Yanhui Bi}
	\altaffiliation{College of Mathematics and Information Science,
		Nanchang Hangkong University}
	
	\author{Bo Yuan}
	\affiliation{College of Mathematics and Information Science,
		Nanchang Hangkong University}
	
	\author{Yuqi Ruan}
	\affiliation{College of Mathematics and Information Science,
		Nanchang Hangkong University}
	
	\author{Tao Zhang}
	\address{School of Mathematics and Statistics,
		Henan Normal University}
	
	\email{biyanhui0523@163.com\quad yuanbo010806@163.com\quad ruanyuqi1023@163.com\quad zhangtao@htu.edu.cn}
	

	\begin{abstract}
		In this paper, we investigate two non-isospectral problems on the loop algebra of the Lie superalgebra $\mathfrak{osp}(1,6)$, and construct two super-integrable systems and their super Hamiltonian structure using the supertrace identity. The resulting super-integrable system can be reduced to the super-AKNS hierarchy under certain conditions. By reconsidering a new $(2+1)$-dimensional non-isospectral problem with spectral matrices satisfying these conditions, we obtain a $(2+1)$-dimensional generalization of the super-AKNS hierarchy.
	\end{abstract}
	
	\maketitle
	
	\section{\label{sec1}Introduction}
	The construction of new isospectral and non-isospectral integrable systems has long been a topic of fundamental importance in soliton theory and the study of integrable systems. Existing results demonstrate that the compatibility condition of spectral problems provides an effective method for constructing such systems. From a mathematical perspective, implementing this approach within the framework of Lie (super)algebras is widely favored, though numerous results have also been obtained in other (super) spaces \cite{ssp,22ssp}.
	
	G.Z. Tu first proposed the trace identity for constructing Hamiltonian structures of continuous and discrete integrable systems \cite{tr,discrete}. Building on this, W.X. Ma introduced the supertrace identity to construct super-Hamiltonian structures for super-integrable systems associated with Lie superalgebras, demonstrating its application to the Lie superalgebra $B(0,1)$ \cite{str}. On $B(0,1)$, he further derived the super-AKNS hierarchy and constructed its corresponding super-Hamiltonian structure. Similarly, working on $\mathfrak{sl}(2,\mathbb{R})$, M. G\"{u}rses and \"{O}. O\u{g}uz also derived the super-AKNS hierarchy \cite{sl2r}. H. F. Wang, Y. F. Zhang and C. Z. Li proposed a method of generation of multi-component super-integrable hierarchies by constructing a new type of multi-component Lie superalgebra $sl(2N,1)$\cite{23H}. On the Lie superalgebra $\mathfrak{sl}(2,1)$, J.W. Han and J. Yu constructed a generalized super-AKNS hierarchy \cite{sl21}. Furthermore, on $\mathfrak{sl}(2,\mathbb{R})$, H.Y. Zhu et al. derived a new integrable generalization of the classical Wadati-Konno-Ichikawa hierarchy \cite{WKI}. Many other scholars have derived notable physical equations on Lie superalgebras, including the super-KdV, super-cKdV, super-KP, and super-Dirac hierarchies, among others \cite{scKdV,sAKNS,twotype,NLS,KP,sDirac,23Dirac,23LS}. The super-GJ and super-Yang hierarchies were presented by S.X. Tao and T.C. Xia \cite{GJYang}, while W.X. Ma and R.G. Zhou provided a soliton hierarchy for a multicomponent AKNS equation \cite{ad}. The general procedure for constructing super-integrable systems and their super-Hamiltonian structures from Lie superalgebras is as follows.
	
	First, consider a spectral problem
	\begin{align}
		\begin{cases}
			\phi_x = U\phi, \\
			\phi_t = V\phi,
		\end{cases}
	\end{align}
	where $U = U(u, \lambda)$, $V = V(v, \lambda) \in \widetilde{\mathfrak{g}}$, and $\widetilde{\mathfrak{g}}$ is a matrix loop algebra of a Lie superalgebra (for a non-isospectral problem, $\lambda_t \neq 0$).
	
	The compatibility condition of the spectral problem yields the zero-curvature equation $U_t - V_x + [U, V] = 0$. Solving the stationary zero-curvature equation $\frac{\partial U}{\partial \lambda} \lambda_t - V_x + [U, V] = 0$ gives a recurrence relation. One may then select a modified term $V^{(n)} = \lambda^n V$, define $V^{(n)}_{+}$ as the part of $V^{(n)}$ containing non-negative powers of $\lambda$, and set $V^{(n)}_{-} = V^{(n)} - V^{(n)}_{+}$. Subsequently, solving the zero-curvature equation
	$\frac{\partial U}{\partial u}u_{t}+\frac{\partial U}{\partial\lambda}\lambda_{t}-V^{(n)}_{+,x}+[U,V^{(n)}_{+}]=0$
	and combining it with the recurrence relation yields a super-integrable hierarchy. Finally, the super-Hamiltonian structure of this hierarchy is constructed via the supertrace identity.
	
	As previously mentioned, scholars have constructed several new super-integrable hierarchies on certain Lie superalgebras. However, research related to $\mathfrak{osp}(l,r)$---one of the four infinite families of classical Lie superalgebras---remains underexplored. In this paper, we choose $U$ and $V$ within the Lie superalgebra $\mathfrak{osp}(1,6)$ and employ the above method to construct a super-integrable system along with its super-Hamiltonian structure. The resulting integrable hierarchy reduces to the super-AKNS hierarchy when specific elements in the spectral matrix are nonzero and all others vanish. By retaining elements at corresponding positions in the spectral matrix, we obtain a new pair of matrices. Based on this pair, we reconsider a new (2+1)-dimensional non-isospectral problem, thereby generalizing the AKNS equations.
	\section{The non-isospectral super-integrable hierarchy of $\mathfrak{osp}(1,6)$ in $(1+1)$-dimension}\label{sec2}
	The orthogonal-symplectic Lie superalgebra $\mathfrak{osp}(l,r)$\quad $(l,r\neq 0)$ is one of the four infinite families of classical Lie superalgebras, defined as follows:\cite{LS}
	\begin{align*}
		\mathfrak{osp}(l,r)_{s}=\left\{X\in\mathfrak{gl}(l,r)_{s}\mid B(Xv,w)+(-1)^{s\lvert v\rvert}B(v,Xw)=0\right\},\quad s\in\mathbb{Z}_{2},
	\end{align*}
	where $B$ is a non-degenerate supersymetric bilinear form and $\lvert v\rvert=\mathrm{deg}(v)$ is the degree of $v$. When $r=0$, $\mathfrak{osp}(l,r)$ reduces to the orthogonal Lie algebra; when $l=0$, it reduces to the symplectic Lie algebra.
	The matrix form of $\mathfrak{osp}(l,r)$ consists of all $(l + r) \times (l + r)$ supermatrices $X$ that satisfy the `infinitesimal invariance' condition: $B(Xv,w)+(-1)^{s\lvert v\rvert}B(v,Xw)=0$. This condition is equivalent to the matrix equation: $X^{st}B+BX=0$, where $X^{st}$ denotes the supertranspose of $X$. The supermatrix $X$ can be expressed in the following block form:
	\begin{align*}
		X=\left(\begin{array}{cc}
			X_{1}&X_{2}\\
			X_{3}&X_{4}
		\end{array}\right),
	\end{align*}
	where $X_{1(l\times l)} ,X_{4(r\times r)} $ are the even part, and $X_{2(l\times r)},X_{3(r\times l)}$ are the odd part. 
	When $l=2m+1,r=2n$, in some basis the matrix of the form $B$ can be written as
	\begin{align*}
		B=\left(\begin{array}{c|c}
			B_{0}&\\
			\hline
			&B_{1}
		\end{array}\right)=
		\left(\begin{array}{ccc|cc}
			0&I_{m}&0&&\\
			I_{m}&0&0&&\\
			0&0&1&&\\
			\hline
			&&&0&I_{n}\\
			&&&-I_{n}&0
		\end{array}\right),
	\end{align*}
	The matrix equation $X^{st}B+BX=0$ is equivalent to the following expressions:
	\begin{align*}
		X_{1}^{t}B_{0}+B_{0}X_{1}=0,\quad  X_{2}^{t}B_{0}+B_{1}X_{3}=0,\quad -X_{3}^{t}B_{1}+B_{0}X_{2}=0, \quad X_{4}^{t}B_{1}+B_{1}X_{4}=0.
	\end{align*}
	Thus, we obtain a matrix in $\mathfrak{osp}(2m+1,2n)$ of the following form:
	\begin{align*}
		\left(\begin{array}{ccc|cc}
			a&b&g&i&l\\
			c&-a^{t}&h&j&m\\
			-h^{t}&-g^{t}&0&k&n\\
			\hline
			m^{t}&l^{t}&n^{t}&d&e\\
			-j^{t}&-i^{t}&-k^{t}&f&-d^{t}
		\end{array}\right),
	\end{align*}
	where $a$ is any $(m\times m)$ matrix, $b$ and $c$ are $(m\times m)$ skew-symmetric matrices, $d$ is any $(n\times n)$ matrix,  $e$ and $f$ are $(n\times n)$ symmetric matrices, $g$ and $h$ are $(m\times 1)$ matrices, $i,j,l,m$ are $(m\times n)$ matrices, $k$ and $n$ are $(1\times n)$ matrices\cite{LS}.
	
	For $m=0, n=3$, the above results yield matrices in $\mathfrak{osp}(1,6)$ of the following form:
	\begin{align*}
		\left(\begin{array}{c|cccccc}
			0&x&y&z&x'&y'&z'\\
			\hline
			-x'&a&d&f&j&m&n\\
			-y'&e&b&h&m&k&o\\
			-z'&g&i&c&n&o&l\\
			x&p&s&t&-a&-e&-g\\
			y&s&q&u&-d&-b&-i\\
			z&t&u&r&-f&-h&-c\\
		\end{array}\right).
	\end{align*}
	Therefore, a basis for $\mathfrak{osp}(1,6)$ is given by
	\begin{align*}
		&E_{1}=e_{22}-e_{55},\ E_{2}=e_{33}-e_{66},\ E_{3}=e_{44}-e_{77},\ E_{4}=e_{23}-e_{65},\ E_{5}=e_{32}-e_{56},\ E_{6}=e_{24}-e_{75},\ E_{7}=e_{42}-e_{57},\\&
		E_{8}=e_{34}-e_{76},\ E_{9}=e_{43}-e_{67},\ E_{10}=e_{25},\ E_{11}=e_{36},\ E_{12}=e_{47},\ E_{13}=e_{26}+e_{35},\ E_{14}=e_{27}+e_{45},\ E_{15}=e_{37}+e_{46},\\&
		E_{16}=e_{52},\ E_{17}=e_{63},\ E_{18}=e_{74},\ E_{19}=e_{53}+e_{62},\ E_{20}=e_{54}+e_{72},\ E_{21}=e_{64}+e_{73},\ E_{22}=e_{12}+e_{51},\ E_{23}=e_{13}+e_{61},\\&
		E_{24}=e_{14}+e_{71},\ E_{25}=e_{15}-e_{21},\ E_{26}=e_{16}-e_{31},\ E_{27}=e_{17}-e_{41},
	\end{align*}
	where $e_{ij}$ is a $(7\times 7)$ matrix with $1$ in the $(i, j)$-th position and $0$ elsewhere. The remaining structure constants are given in Appendix~\ref{appA}, where $\mathfrak{osp}(1,6)=\mathfrak{osp}(1,6)_{\bar{0}}\oplus\mathfrak{osp}(1,6)_{\bar{1}}$, $\mathfrak{osp}(1,6)_{\bar{0}}=\mathrm{span}\left\{E_{1},\dots,E_{21}\right\}$ are even and $\mathfrak{osp}(1,6)_{\bar{1}}=\mathrm{span}\left\{E_{22},\dots,E_{27}\right\}$ are odd, and $\left[\cdot,\cdot\right]$ and $\left[\cdot,\cdot\right]_{+}$ denote the commutator and the anticommutator.
	The loop algebra corresponding to $\mathfrak{osp}(1,6)$ is given by
	\begin{align*}
		\widetilde{\mathfrak{osp}}(1,6)=\mathrm{span}\left\{E_{i}(n)\right\}_{i=1}^{27}.
	\end{align*}
	where $E_{i}(n)=E_{i}\lambda^{n}$. Building on the above, by taking $A=\sum_{i=1}^{27}a_{i}E_{i}$ and $B=\sum_{i=1}^{27}b_{i}E_{i}\in\widetilde{\mathfrak{osp}}(1,6)$, a direct computation yields the Killing form of $\widetilde{\mathfrak{osp}}(1,6)$ as follows:
	\begin{align*}
		\mathrm{str}(\mathrm{ad}_{A}\mathrm{ad}_{B})=-7\mathrm{str}(AB)
	\end{align*}
	where $\mathrm{str}$  denotes the supertrace, and $\mathrm{ad}_{A}$ represents the adjoint action of $A\in\widetilde{\mathfrak{osp}}(1,6)$ on itself, defined as follows:
	\begin{align*}
		\mathrm{ad}_{A}B=[A,B],\quad B\in\widetilde{\mathfrak{osp}}(1,6).
	\end{align*}
	The supertrace and the adjoint action satisfy the symmetric property:
	\begin{align*}
		\mathrm{str}(\mathrm{ad}_{A}\mathrm{ad}_{B})=\mathrm{str}(\mathrm{ad}_{B}\mathrm{ad}_{A})
	\end{align*}
	and the invariance property under the $[\cdot,\cdot]$:
	\begin{align*}
		\mathrm{ad}_{A}\mathrm{ad}_{[B,C]}=\mathrm{ad}_{[A,B]}\mathrm{ad}_{C},
	\end{align*}
	where $A,B,C\in\widetilde{\mathfrak{osp}}(1,6)$.
	Let $U,V\in\widetilde{\mathfrak{osp}}(1,6)$,
	\begin{align*}
		U=\left(\begin{array}{c|cccccc}
			0&u_{13}&u_{14}&u_{15}&u_{16}&u_{17}&u_{18}\\
			\hline
			-u_{16}&\lambda&0&0&u_{1}&u_{4}&u_{5}\\
			-u_{17}&0&\lambda&0&u_{4}&u_{2}&u_{6}\\
			-u_{18}&0&0&\lambda&u_{5}&u_{6}&u_{3}\\
			u_{13}&u_{7}&u_{10}&u_{11}&-\lambda&0&0\\
			u_{14}&u_{10}&u_{8}&u_{12}&0&-\lambda&0\\
			u_{15}&u_{11}&u_{12}&u_{9}&0&0&-\lambda
		\end{array}\right)=\left(\begin{array}{cc}
			0&U_{1}\\
			U_{2}&U_{3}
		\end{array}\right)
	\end{align*}
	and
	\begin{align*}
		V=\left(\begin{array}{c|cccccc}
			0&\alpha&\beta&\gamma&\alpha'&\beta'&\gamma'\\
			\hline
			-\alpha'&a&d&f&k&n&o\\
			-\beta'&e&b&h&n&l&p\\
			-\gamma'&g&j&c&o&p&m\\
			\alpha&q&v&w&-a&-e&-g\\
			\beta&v&r&y&-d&-b&-j\\
			\gamma&w&y&s&-f&-h&-c
		\end{array}\right)=\sum_{i\geq0}
		\left(\begin{array}{c|cccccc}
			0&\alpha_{i}&\beta_{i}&\gamma_{i}&\alpha'_{i}&\beta'_{i}&\gamma'_{i}\\
			\hline
			-\alpha'_{i}&a_{i}&d_{i}&f_{i}&k_{i}&n_{i}&o_{i}\\
			-\beta'_{i}&e_{i}&b_{i}&h_{i}&n_{i}&l_{i}&p_{i}\\
			-\gamma'_{i}&g_{i}&j_{i}&c_{i}&o_{i}&p_{i}&m_{i}\\
			\alpha_{i}&q_{i}&v_{i}&w_{i}&-a_{i}&-e_{i}&-g_{i}\\
			\beta_{i}&v_{i}&r_{i}&y_{i}&-d_{i}&-b_{i}&-j_{i}\\
			\gamma_{i}&w_{i}&y_{i}&s_{i}&-f_{i}&-h_{i}&-c_{i}
		\end{array}\right)\lambda^{-i}=\left(\begin{array}{cc}
			0&V_{1}\\
			V_{1}&V_{3}
		\end{array}\right),
	\end{align*}
	where $u_{1},\dots,u_{12}$ are even variables, $u_{13},\dots,u_{18}$ are odd variables, and $a,\dots,y$ belong to a commutative field, $\alpha,\dots,\gamma'$ belong to an anticommutative field,
	\begin{align*}
		[U,V]_{s}=\left(\begin{array}{cc}
			U_{1}V_{2}+V_{1}U_{2}&U_{1}V_{3}-V_{1}U_{3}\\
			U_{3}V_{2}-V_{3}U_{2}&U_{2}V_{1}+V_{2}U_{1}+U_{3}V_{3}-V_{3}U_{3}
		\end{array}\right).
	\end{align*}
	Consider a non-isospectral problem
	\begin{align*}
		\begin{cases}
			\phi_{x}=U\phi\\
			\phi_{t}=V\phi\\
			\lambda_{t}=\sum_{i\geq0}z_{i}(t)\lambda^{-i}
		\end{cases}
	\end{align*}
	The stationary zero curvature representation $V_{x}=\frac{\partial U}{\partial\lambda}\lambda_{t}+[U,V]$ gives
	
	\begin{align}
		\begin{cases}
			a_{x}=u_{1}q+u_{4}v+u_{5}w-u_{7}k-u_{10}n-u_{11}o-u_{13}\alpha'-u_{16}\alpha+z(t),\\
			b_{x}=u_{2}r+u_{4}v+u_{6}y-u_{8}l-u_{10}n-u_{12}p-u_{14}\beta'-u_{17}\beta+z(t),\\
			c_{x}=u_{3}s+u_{5}w+u_{6}y-u_{9}m-u_{11}o-u_{12}p-u_{15}\gamma'-u_{18}\gamma+z(t),\\
			d_{x}=u_{1}v+u_{4}r+u_{5}y-u_{8}n-u_{10}k-u_{12}o-u_{14}\alpha'-u_{16}\beta,\\
			e_{x}=u_{2}v+u_{4}q+u_{6}w-u_{7}n-u_{10}l-u_{11}p-u_{13}\beta'-u_{17}\alpha,\\
			f_{x}=u_{1}w+u_{4}y+u_{5}s-u_{9}o-u_{11}k-u_{12}n-u_{15}\alpha'-u_{16}\gamma,\\
			g_{x}=u_{3}w+u_{5}q+u_{6}v-u_{7}o-u_{10}p-u_{11}m-u_{13}\gamma'-u_{18}\alpha,\\
			h_{x}=u_{2}y+u_{4}w+u_{6}s-u_{9}p-u_{11}n-u_{12}l-u_{15}\beta'-u_{17}\gamma,\\
			j_{x}=u_{3}y+u_{5}v+u_{6}r-u_{8}p-u_{10}o-u_{12}m-u_{14}\gamma'-u_{18}\beta,\\
			k_{x}=2\lambda k-2u_{1}a-2u_{4}d-2u_{5}f-2u_{16}\alpha',\\
			l_{x}=2\lambda l-2u_{2}b-2u_{4}e-2u_{6}h-2u_{17}\beta',\\
			m_{x}=2\lambda m-2u_{3}c-2u_{5}g-2u_{6}j-2u_{18}\gamma',\\
			n_{x}=2\lambda n-u_{1}e-u_{2}d-u_{4}a-u_{4}b-u_{5}h-u_{6}f-u_{16}\beta'-u_{17}\alpha',\\
			o_{x}=2\lambda o-u_{1}g-u_{3}f-u_{4}j-u_{5}a-u_{5}c-u_{6}d-u_{16}\gamma'-u_{18}\alpha',\\
			p_{x}=2\lambda p-u_{2}j-u_{3}h-u_{4}g-u_{5}e-u_{6}b-u_{6}c-u_{17}\gamma'-u_{18}\beta'\\
			q_{x}=-2\lambda q+2u_{7}a+2u_{10}e+2u_{11}g+2u_{13}\alpha,\\
			r_{x}=-2\lambda r+2u_{8}b+2u_{10}d+2u_{12}j+2u_{14}\beta,\\
			s_{x}=-2\lambda s+2u_{9}c+2u_{11}f+2u_{12}h+2u_{15}\gamma,\\
			v_{x}=-2\lambda v+u_{7}d+u_{8}e+u_{10}a+u_{10}b+u_{11}j+u_{12}g+u_{13}\beta+u_{14}\alpha,\\
			w_{x}=-2\lambda w+u_{7}f+u_{9}g+u_{10}h+u_{11}a+u_{11}c+u_{12}e+u_{13}\gamma+u_{15}\alpha,\\
			y_{x}=-2\lambda y+u_{8}h+u_{9}j+u_{10}f+u_{11}d+u_{12}b+u_{12}c+u_{14}\gamma+u_{15}\beta,\\
			\alpha_{x}=-\lambda\alpha-u_{7}\alpha'-u_{10}\beta'-u_{11}\gamma'+u_{13}a+u_{14}e+u_{15}g+u_{16}q+u_{17}v+u_{18}w,\\
			\beta_{x}=-\lambda\beta-u_{8}\beta'-u_{10}\alpha'-u_{12}\gamma'+u_{13}d+u_{14}b+u_{15}j+u_{16}v+u_{17}r+u_{18}y,\\
			\gamma_{x}=-\lambda\gamma-u_{9}\gamma'-u_{11}\alpha'-u_{12}\beta'+u_{13}f+u_{14}h+u_{15}c+u_{16}w+u_{17}y+u_{18}s,\\
			\alpha'_{x}=\lambda\alpha'-u_{1}\alpha-u_{4}\beta-u_{5}\gamma+u_{13}k+u_{14}n+u_{15}o-u_{16}a-u_{17}d-u_{18}f,\\
			\beta'_{x}=\lambda\beta'-u_{2}\beta-u_{4}\alpha-u_{6}\gamma+u_{13}n+u_{14}l+u_{15}p-u_{16}e-u_{17}b-u_{18}h,\\
			\gamma'_{x}=\lambda\gamma'-u_{3}\gamma-u_{5}\alpha-u_{6}\beta+u_{13}o+u_{14}p+u_{15}m-u_{16}g-u_{17}j-u_{18}c.
		\end{cases}\label{r1}
	\end{align}
	
	Take the initial values
	\begin{align*}
		a_{0}=b_{0}=c_{0}=1,\ d_{0}=e_{0}=\dots=w_{0}=y_{0}=\alpha_{0}=\dots=\gamma'_{0}=z_{0}(t)=0.
	\end{align*}
	we have
	\begin{align}
		a_{1}=&z_{1}(t)x+\partial^{-1}(-2u_{13}u_{16}),\ b_{1}=z_{1}(t)x+\partial^{-1}(-2u_{14}u_{17}),\ c_{1}=z_{1}(t)x+\partial^{-1}(-2u_{15}u_{18}),\ d_{1}=\partial^{-1}(-2u_{14}u_{16}),\nonumber\\ e_{1}=&\partial^{-1}(-2u_{13}u_{17}),\  f_{1}=\partial^{-1}(-2u_{15}u_{16}),\ g_{1}=\partial^{-1}(-2u_{13}u_{18}),\ h_{1}=\partial^{-1}(-2u_{15}u_{17}),\ j_{1}=\partial^{-1}(-2u_{14}u_{18}),\ k_{1}=u_{1},\nonumber\\ l_{1}=&u_{2}, m_{1}=u_{3},\ n_{1}=u_{4},\ o_{1}=u_{5},\ p_{1}=u_{6},\ q_{1}=u_{7},\ r_{1}=u_{8},\ s_{1}=u_{9},\ v_{1}=u_{10},\ w_{1}=u_{11},\ y_{1}=u_{12},\ \alpha_{1}=u_{13},\ \beta_{1}=u_{14},\nonumber\\ \gamma_{1}=&u_{15},\ \alpha'_{1}=u_{16},\ \beta'_{1}=u_{17},\ \gamma'_{1}=u_{18},\ 
		k_{2}=\frac{1}{2}u_{1x}+u_{1}z_{1}(t)x+u_{1}\partial^{-1}(-2u_{13}u_{16})+u_{4}\partial^{-1}(-2u_{14}u_{16})+u_{5}\partial^{-1}(-2u_{15}u_{16})+u_{16}^{2},\nonumber\\ 
		l_{2}=&\frac{1}{2}u_{2x}+u_{2}z_{1}(t)x+u_{2}\partial^{-1}(-2u_{14}u_{17})+u_{4}\partial^{-1}(-2u_{13}u_{17})+u_{6}\partial^{-1}(-2u_{15}u_{17})+u_{17}^{2},\nonumber\\ 
		m_{2}=&\frac{1}{2}u_{3x}+u_{3}z_{1}(t)x+u_{3}\partial^{-1}(-2u_{15}u_{18})+u_{5}\partial^{-1}(-2u_{13}u_{18})+u_{6}\partial^{-1}(-2u_{14}u_{18})+u_{18}^{2},\nonumber\\ 
		n_{2}=&\frac{1}{2}u_{4x}+\frac{1}{2}u_{1}\partial^{-1}(-2u_{13}u_{17})+\frac{1}{2}u_{2}\partial^{-1}(-2u_{14}u_{16})+u_{4}z_{1}(t)x+\frac{1}{2}u_{4}\partial^{-1}(-2u_{13}u_{16})+\frac{1}{2}u_{4}\partial^{-1}(-2u_{14}u_{17})\nonumber\\&+\frac{1}{2}u_{6}\partial^{-1}(-2u_{15}u_{16})+u_{16}u_{17},\  
		o_{2}=\frac{1}{2}u_{5x}+\frac{1}{2}u_{1}\partial^{-1}(-2u_{13}u_{18})+\frac{1}{2}u_{3}\partial^{-1}(-2u_{15}u_{16})+\frac{1}{2}u_{4}\partial^{-1}(-2u_{14}u_{18})+u_{5}z_{1}(t)x\nonumber\\&+\frac{1}{2}u_{5}\partial^{-1}(-2u_{13}u_{16})+\frac{1}{2}u_{5}\partial^{-1}(-2u_{15}u_{18})+\frac{1}{2}u_{6}\partial^{-1}(-2u_{14}u_{16})+u_{16}u_{18},\nonumber\\ p_{2}=&\frac{1}{2}u_{6x}+\frac{1}{2}u_{2}\partial^{-1}(-2u_{15}u_{16})+\frac{1}{2}u_{3}\partial^{-1}(-2u_{15}u_{17})+\frac{1}{2}u_{4}\partial^{-1}(-2u_{13}u_{18})+\frac{1}{2}u_{5}\partial^{-1}(-2u_{13}u_{17})+u_{6}z_{1}(t)x\nonumber\\&+\frac{1}{2}u_{6}\partial^{-1}(-2u_{14}u_{17})+\frac{1}{2}u_{6}\partial^{-1}(-2u_{15}u_{18})+u_{17}u_{18},\ q_{2}=-\frac{1}{2}u_{7x}+u_{7}z_{1}(t)x+u_{7}\partial^{-1}(-2u_{13}u_{16})+u_{10}\partial^{-1}(-2u_{13}u_{17})\nonumber\\&+u_{11}\partial^{-1}(-2u_{13}u_{18})+u_{13}^{2},\ r_{2}=-\frac{1}{2}u_{8x}+u_{8}z_{1}(t)x+u_{8}\partial^{-1}(-2u_{14}u_{17})+u_{10}\partial^{-1}(-2u_{14}u_{16})+u_{12}\partial^{-1}(-2u_{14}u_{18})+u_{14}^{2},\nonumber\\ s_{2}=&-\frac{1}{2}u_{9x}+u_{9}z_{1}(t)x+u_{9}\partial^{-1}(-2u_{15}u_{18})+u_{11}\partial^{-1}(-2u_{15}u_{16})+u_{12}\partial^{-1}(-2u_{15}u_{17})+u_{15}^{2},\label{value} \\
		v_{2}=&-\frac{1}{2}u_{10x}+\frac{1}{2}u_{7}\partial^{-1}(-2u_{14}u_{16})+\frac{1}{2}u_{8}\partial^{-1}(-2u_{13}u_{17})+u_{10}z_{1}(t)x+\frac{1}{2}u_{10}\partial^{-1}(-2u_{13}u_{16})+\frac{1}{2}u_{10}\partial^{-1}(-2u_{14}u_{17})\nonumber\\&+\frac{1}{2}u_{11}\partial^{-1}(-2u_{14}u_{18})+\frac{1}{2}u_{12}\partial^{-1}(-2u_{13}u_{18})+u_{13}u_{14},\   
		w_{2}=-\frac{1}{2}u_{11x}+\frac{1}{2}u_{7}\partial^{-1}(-2u_{15}u_{16})+\frac{1}{2}u_{9}\partial^{-1}(-2u_{13}u_{18})\nonumber\\&+\frac{1}{2}u_{10}\partial^{-1}(-2u_{15}u_{17})+u_{11}z_{1}(t)x+\frac{1}{2}u_{11}\partial^{-1}(-2u_{13}u_{16})+\frac{1}{2}u_{11}\partial^{-1}(-2u_{15}u_{18})+\frac{1}{2}u_{12}\partial^{-1}(-2u_{13}u_{17})+u_{13}u_{15},\nonumber\\ 
		y_{2}=&-\frac{1}{2}u_{12x}+\frac{1}{2}u_{8}\partial^{-1}(-2u_{15}u_{17})+\frac{1}{2}u_{9}\partial^{-1}(-2u_{14}u_{18})+\frac{1}{2}u_{10}\partial^{-1}(-2u_{15}u_{16})+\frac{1}{2}u_{11}\partial^{-1}(-2u_{14}u_{16})+u_{12}z_{1}(t)x\nonumber\\&+\frac{1}{2}u_{12}\partial^{-1}(-2u_{14}u_{17})+\frac{1}{2}u_{12}\partial^{-1}(-2u_{15}u_{18})+u_{14}u_{15},\ \alpha_{2}=-u_{13x}+u_{13}z_{1}(t)x+u_{13}\partial^{-1}(-2u_{13}u_{16})+u_{14}\partial^{-1}(-2u_{13}u_{17}),\nonumber\\ \beta_{2}=&-u_{14x}+u_{14}z_{1}(t)x+u_{14}\partial^{-1}(-2u_{14}u_{17})+u_{15}\partial^{-1}(-2u_{14}u_{18}),\nonumber\\ \gamma_{2}=&-u_{15x}+u_{13}\partial^{-1}(-2u_{15}u_{16})+u_{14}\partial^{-1}(-2u_{15}u_{17})+u_{15}z_{1}(t)x+u_{15}\partial^{-1}(-2u_{15}u_{18}),\nonumber\\ \alpha'_{2}=&u_{16x}+u_{16}z_{1}(t)x+u_{16}\partial^{-1}(-2u_{13}u_{16})+u_{17}\partial^{-1}(-2u_{14}u_{16})+u_{18}\partial^{-1}(-2u_{15}u_{16}),\nonumber\\ \beta'_{2}=&u_{17x}+u_{16}\partial^{-1}(-2u_{13}u_{17})+u_{17}z_{1}(t)x+u_{17}\partial^{-1}(-2u_{14}u_{17})+u_{18}\partial^{-1}(-2u_{15}u_{17}),\nonumber\\ \gamma'_{2}=&u_{18x}+u_{16}\partial^{-1}(-2u_{13}u_{18})+u_{17}\partial^{-1}(-2u_{14}u_{18})+u_{18}z_{1}(t)x+u_{18}\partial^{-1}(-2u_{15}u_{18}),\ \cdots\cdots\nonumber
	\end{align}
	Now, taking
	\begin{align*}
		V^{(n)}=\lambda^{n}V=\sum_{i\geq0}(a_{i},\dots,y_{i},\dots,\alpha_{i},\dots,\gamma'_{i})^{t}\lambda^{n-i},\quad V_{+}^{(n)}=\sum_{i=0}^{n}(a_{i},\dots,y_{i},\dots,\alpha_{i},\dots,\gamma'_{i})^{t}\lambda^{n-i},\quad V_{-}^{(n)}=V^{(n)}-V_{+}^{(n)},
	\end{align*}
	then the zero curvature equation $\frac{\partial U}{\partial u}u_{t}+\frac{\partial U}{\partial\lambda}\lambda_{t}-V_{+x}^{n}+[U,V_{+}^{n}]=0$ leads to the following super-integrable hierarchy
	\begin{align}
		u_{t_{n}}=&\left(u_{1},u_{2},u_{3},u_{4},u_{5},u_{6},u_{7},u_{8},u_{9},u_{10},u_{11},u_{12},u_{13},u_{14},u_{15},u_{16},u_{17},u_{18}\right)^{t}_{t_{n}}\notag\\
		=&(2k_{n+1},2l_{n+1},2m_{n+1},2n_{n+1},2o_{n+1},2p_{n+1},-2q_{n+1},-2r_{n+1},-2s_{n+1},\notag\\&-2v_{n+1},-2w_{n+1},-2y_{n+1},-\alpha_{n+1},-\beta_{n+1},-\gamma_{n+1},\alpha'_{n+1},\beta'_{n+1},\gamma'_{n+1})^{t}\notag\\
		=&J_{1}(-q_{n+1},-r_{n+1},-s_{n+1},-2v_{n+1},-2w_{n+1},-2y_{n+1},-k_{n+1},-l_{n+1},-m_{n+1},\notag\\&-2n_{n+1},-2o_{n+1},-2p_{n+1},2\alpha'_{n+1},2\beta'_{n+1},2\gamma'_{n+1},-2\alpha_{n+1},-2\beta_{n+1},-2\gamma_{n+1})^{t}\notag\\=&J_{1}P_{1,n+1},\label{h1}
	\end{align}
	where $J_{1}$ is
	\begin{align*}
		J_{1}=\left(\begin{array}{cccccccccccc|cccccc}
			0&0&0&0&0&0&-2&0&0&0&0&0&0&0&0&0&0&0\\
			0&0&0&0&0&0&0&-2&0&0&0&0&0&0&0&0&0&0\\
			0&0&0&0&0&0&0&0&-2&0&0&0&0&0&0&0&0&0\\
			0&0&0&0&0&0&0&0&0&-1&0&0&0&0&0&0&0&0\\
			0&0&0&0&0&0&0&0&0&0&-1&0&0&0&0&0&0&0\\
			0&0&0&0&0&0&0&0&0&0&0&-1&0&0&0&0&0&0\\
			2&0&0&0&0&0&0&0&0&0&0&0&0&0&0&0&0&0\\
			0&2&0&0&0&0&0&0&0&0&0&0&0&0&0&0&0&0\\
			0&0&2&0&0&0&0&0&0&0&0&0&0&0&0&0&0&0\\
			0&0&0&1&0&0&0&0&0&0&0&0&0&0&0&0&0&0\\
			0&0&0&0&1&0&0&0&0&0&0&0&0&0&0&0&0&0\\
			0&0&0&0&0&1&0&0&0&0&0&0&0&0&0&0&0&0\\
			\hline
			0&0&0&0&0&0&0&0&0&0&0&0&0&0&0&\frac{1}{2}&0&0\\
			0&0&0&0&0&0&0&0&0&0&0&0&0&0&0&0&\frac{1}{2}&0\\
			0&0&0&0&0&0&0&0&0&0&0&0&0&0&0&0&0&\frac{1}{2}\\
			0&0&0&0&0&0&0&0&0&0&0&0&\frac{1}{2}&0&0&0&0&0\\
			0&0&0&0&0&0&0&0&0&0&0&0&0&\frac{1}{2}&0&0&0&0\\
			0&0&0&0&0&0&0&0&0&0&0&0&0&0&\frac{1}{2}&0&0&0\\
		\end{array}\right).
	\end{align*}
	
	From the recurrence relations (\ref{r1}), we have
	\begin{align*}
		P_{1,n+1}=L_{1}P_{1,n}+u_{L}z_{n}(t)x
	\end{align*}
	where
	\begin{align*}
		u_{L}=(-u_{7},-u_{8},-u_{9},-2u_{10},-2u_{11},-2u_{12},-u_{1},-u_{2},-u_{3},-2u_{4},-2u_{5},-2u_{6},2u_{16},2u_{17},2u_{18},-2u_{13},-2u_{14},-2u_{15})^{t}
	\end{align*} and $L_{1}$ is given in Appendix~\ref{appB}.
	We derive the super-Hamiltonian structure of (\ref{h1}) via the supertrace identity\cite{str}
	\begin{align*}
		&\left\langle V,\frac{\partial U}{\partial\lambda}\right\rangle = -2a-2b-2c ,\
		\left\langle V,\frac{\partial U}{\partial u_{1}}\right\rangle = -q ,\
		\left\langle V,\frac{\partial U}{\partial u_{2}}\right\rangle = -r ,\
		\left\langle V,\frac{\partial U}{\partial u_{3}}\right\rangle = -s ,\\
		&\left\langle V,\frac{\partial U}{\partial u_{4}}\right\rangle = -2v ,\
		\left\langle V,\frac{\partial U}{\partial u_{5}}\right\rangle = -2w ,\
		\left\langle V,\frac{\partial U}{\partial u_{6}}\right\rangle = -2y ,\
		\left\langle V,\frac{\partial U}{\partial u_{7}}\right\rangle = -k ,\
		\left\langle V,\frac{\partial U}{\partial u_{8}}\right\rangle = -l ,\\
		&\left\langle V,\frac{\partial U}{\partial u_{9}}\right\rangle = -m ,\
		\left\langle V,\frac{\partial U}{\partial u_{10}}\right\rangle = -2n ,\
		\left\langle V,\frac{\partial U}{\partial u_{11}}\right\rangle = -2o ,\
		\left\langle V,\frac{\partial U}{\partial u_{12}}\right\rangle = -2p .\
		\left\langle V,\frac{\partial U}{\partial u_{13}}\right\rangle = 2\alpha' ,\\
		&\left\langle V,\frac{\partial U}{\partial u_{14}}\right\rangle = 2\beta' ,\
		\left\langle V,\frac{\partial U}{\partial u_{15}}\right\rangle = 2\gamma' ,\
		\left\langle V,\frac{\partial U}{\partial u_{16}}\right\rangle = -2\alpha ,\
		\left\langle V,\frac{\partial U}{\partial u_{17}}\right\rangle = -2\beta ,\
		\left\langle V,\frac{\partial U}{\partial u_{18}}\right\rangle = -2\gamma ,\\
	\end{align*}
	where $	\left\langle X,Y\right\rangle=\mathrm{str}(XY)$. According to the Killing form $\mathrm{str}(\mathrm{ad}_{A}\mathrm{ad}_{B})=-7\mathrm{str}(AB)$, we have
	\begin{align*}
		&\mathrm{str}(\mathrm{ad}_{V}\mathrm{ad}_{\frac{\partial U}{\partial\lambda}})=14a+14b+14c,\\ &\mathrm{str}(\mathrm{ad}_{V}\mathrm{ad}_{\frac{\partial U}{\partial u}})=(7q,7r,7s,14v,14w,14y,7k,7l,7m,14n,14o,14p,-14\alpha',-14\beta',-14\gamma',14\alpha,14\beta,14\gamma)^{t}
	\end{align*}
	Substituting the above formulate into the supertrace identity\cite{str} \begin{align*}
		\frac{\delta}{\delta u}\int\mathrm{str}(\mathrm{ad}_{V}\mathrm{ad}_{\frac{\partial U}{\partial\lambda}})dx=\lambda^{-\tau}\frac{\partial}{\partial\lambda}\lambda^{\tau}\mathrm{str}(\mathrm{ad}_{V}\mathrm{ad}_{\frac{\partial U}{\partial u}})
	\end{align*}
	and balancing coefficients of each power of $\lambda$ in the above equality gives rise to
	\begin{align*}
		&\frac{\delta}{\delta u}(14a_{n+1}14b_{n+1}14c_{n+1})\\&=(\tau-n)
		(7q_{n},7r_{n},7s_{n},14v_{n},14w_{n},14y_{n},7k_{n},7l_{n},7m_{n},
		14n_{n},14o_{n},14p_{n},-14\alpha'_{n},-14\beta'_{n},-14\gamma'_{n},14\alpha_{n},14\beta_{n},14\gamma_{n})^{t}.
	\end{align*}
	Taking $n=1$ and using the above equation together with equation (\ref{value}), we obtain $\tau=0$. Therefore, the expression for $P_{n+1}$ above can be written in the following form:
	\begin{align*}
		P_{1,n+1}=&(-q_{n+1},-r_{n+1},-s_{n+1},-2v_{n+1},-2w_{n+1},-2y_{n+1},-k_{n+1},-l_{n+1},-m_{n+1},\\&-2n_{n+1},-2o_{n+1},-2p_{n+1},2\alpha'_{n+1},2\beta'_{n+1},2\gamma'_{n+1},-2\alpha_{n+1},-2\beta_{n+1},-2\gamma_{n+1})^{t}\\=&\frac{\delta}{\delta u}(\frac{2}{n+1}(a_{n+2}+b_{n+2}+c_{n+2})),
	\end{align*}
	consequently, the hierarchy (\ref{h1}) can be expressed in the following form:
	\begin{align*}
		u_{t,n}=J_{1}P_{1,n+1}=J_{1}\frac{\delta H_{n+1}^{1}}{\delta u}=J_{1}L_{1}\frac{\delta H_{n}^{1}}{\delta u}+J_{1}u_{L}z_{n}(t)x,\ H_{n+1}^{1}=(\frac{2}{n+1})(a_{n+2}+b_{n+2}+c_{n+2}),\ n\geq0.
	\end{align*}
	Thus, we have constructed a super‑integrable system on $\mathfrak{osp}(1,6)$. Furthermore, by examining the matrix $J_{1}$, the following conclusion can be drawn:
	when one set of $\{u_{1},u_{7}\},\{u_{2},u_{8}\},\{u_{3},u_{9}\}$ is nonzero and the others vanish, the hierarchy (\ref{h1}) reduces to the AKNS hierarchy\cite{tr}
	\begin{align}
		u_{t_{n}}=\left(\begin{array}{c}
			u_{1}\\u_{7}
		\end{array}\right)=\left(\begin{array}{c}
			2k_{n+1}\\-2q_{n+1}
		\end{array}\right)=\left(\begin{array}{cc}
			0&-2\\2&0
		\end{array}\right)\left(\begin{array}{c}
			-q_{n+1}\\-k_{n+1}
		\end{array}\right)\label{AKNS}
	\end{align}
	When one set from $\{u_{1},u_{7}\},\{u_{2},u_{8}\},\{u_{3},u_{9}\}$ and one set from $\{u_{13},u_{16}\},\{u_{14},u_{17}\},\{u_{15},u_{18}\}$ are nonzero while all others vanish, the hierarchy (\ref{h1}) reduces to the super-AKNS hierarchy
	\begin{align}
		u_{t_{n}}=\left(\begin{array}{c}
			u_{1}\\u_{7}\\u_{13}\\u_{16}
		\end{array}\right)=\left(\begin{array}{c}
			2k_{n+1}\\-2q_{n+1}\\-\alpha_{n+1}\\\alpha'_{n+1}
		\end{array}\right)=\left(\begin{array}{cccc}
			0&-2&0&0\\2&0&0&0\\0&0&0&\frac{1}{2}\\0&0&\frac{1}{2}&0
		\end{array}\right)\left(\begin{array}{c}
			-q_{n+1}\\-k_{n+1}\\2\alpha'_{n+1}\\-2\alpha_{n+1}
		\end{array}\right)\label{sAKNS}
	\end{align}
	\section{A generalized super-AKNS hierarchy in $(2+1)$ dimension}\label{sec3}
	When $u_{1},u_{7},u_{13},u_{16}$ are nonzero and all other elements in $U$ vanish, the hierarchy (\ref{h1}) reduces to the super-AKNS hierarchy. In this case, we rewrite $U$ and $V$ in the following form:
	\begin{align*}
		U_{sAKNS}=\left(\begin{array}{c|cc}
			0&u_{3}&u_{4}\\
			\hline
			-u_{4}&\lambda&u_{1}\\
			u_{3}&u_{2}&-\lambda
		\end{array}\right),\quad V_{sAKNS}=\left(\begin{array}{c|cc}
			0&d&e\\
			\hline
			-e&a&b\\
			d&c&-a
		\end{array}\right)=\sum_{i\geq0}\left(\begin{array}{c|cc}
			0&d_{i}&e_{i}\\
			\hline
			-e_{i}&a_{i}&b_{i}\\
			d_{i}&c_{i}&-a_{i}
		\end{array}\right)\lambda^{-i},
	\end{align*}
	where $U_{sAKNS},V_{sAKNS}\in\widetilde{\mathfrak{osp}}(1,2)$. Let
	\begin{align*}
		\frac{\partial}{\partial\upsilon}=\frac{\partial}{\partial y}-\frac{\partial}{\partial x},\quad \frac{\partial}{\partial\omega}=\frac{\partial}{\partial t}-\frac{\partial}{\partial x}
	\end{align*}
	where $x$ and $y$ are spatial variables, and $t$ is the temporal variable.
	Consider a non-isospectral problem
	\begin{align*}
		\begin{cases}
			\phi_{\upsilon}=U_{sAKNS}\phi\\
			\phi_{\omega}=V_{sAKNS}\phi\\
			\lambda_{t}=\sum_{i\geq0}z_{i}(t)\lambda^{-i}
		\end{cases}
	\end{align*}
	In the following, $U_{sAKNS}$ and $V_{sAKNS}$ will be denoted succinctly as $U$ and $V$, respectively.
	The zero-curvature equation and the stationary zero-curvature equation can be rewritten in the following forms, respectively
	\begin{align*}
		U_{t}-U_{x}+V_{x}-V_{y}+[U,V]_{s}=0,\quad V_{y}-V_{x}=\frac{\partial U}{\partial\lambda}\lambda_{t}+[U,V]_{s}
	\end{align*}
	and satisfies the compatibility condition for the following Lax pair
	\begin{align*}
		\begin{cases}
			\phi_{y}=\phi_{x}+U\phi\\
			\phi_{t}=\phi_{x}+V\phi.
		\end{cases}
	\end{align*}
	The stationary zero curvature representation $V_{y}-V_{x}=\frac{\partial U}{\partial\lambda}\lambda_{t}+[U,V]_{s}$ gives
	\begin{align}
		\begin{cases}
			a_{y}-a_{x}=u_{1}c-u_{2}b-u_{3}e-u_{4}d+z(t)\\
			b_{y}-b_{x}=2\lambda b-2u_{1}a-2u_{4}e\\
			c_{y}-c_{x}=-2\lambda c+2u_{2}a+2u_{3}d\\
			d_{y}-d_{x}=-\lambda d-u_{2}e+u_{3}a+u_{4}c\\
			e_{y}-e_{x}=\lambda e-u_{1}d+u_{3}b-u_{4}a
		\end{cases}\label{r2}
	\end{align}
	Take the initial values
	\begin{align*}
		a_{0}=1,\; b_{0}=c_{0}=d_{0}=e_{0}=z_{0}(t)=0
	\end{align*}
	we can compute the leading terms as follows:
	\begin{align*}
		a_{1}=&\partial_{x}^{-1}(a_{1y}+2u_{3}u_{4})-z_{1}(t)x,\quad b_{1}=u_{1},\quad c_{1}=u_{2},\quad d_{1}=u_{3},\quad e_{1}=u_{4},\\ b_{2}=&\frac{1}{2}(u_{1y}-u_{1x})+u_{1}\partial_{x}^{-1}(a_{1y}+2u_{3}u_{4})-u_{1}z_{1}(t)x+u_{4}^{2},\quad c_{2}=-\frac{1}{2}(u_{2y}-u_{2x})+u_{2}\partial_{x}^{-1}(a_{1y}+2u_{3}u_{4})-u_{2}z_{1}(t)x+u_{3}^{2},\\
		d_{2}=&-(u_{3y}-u_{3x})+u_{3}\partial_{x}^{-1}(a_{1y}+2u_{3}u_{4})-u_{3}z_{1}(t)x,\quad e_{2}=(u_{4y}-u_{4x})+u_{4}\partial_{x}^{-1}(a_{1y}+2u_{3}u_{4})-u_{4}z_{1}(t)x\\
		&\cdots\cdots
	\end{align*}
	Taking
	\begin{align*}
		V^{n}=\lambda^{n}V=\sum_{i\geq0}\left(\begin{array}{ccc}
			0&d_{i}&e_{i}\\
			-e_{i}&a_{i}&b_{i}\\
			d_{i}&c_{i}&-a_{i}
		\end{array}\right)\lambda^{n-i},\quad V^{n}_{+}=\sum_{i=0}^{n}\left(\begin{array}{ccc}
			0&d_{i}&e_{i}\\
			-e_{i}&a_{i}&b_{i}\\
			d_{i}&c_{i}&-a_{i}
		\end{array}\right)\lambda^{n-i},\quad V^{n}_{-}=V^{n}-V^{n}_{+}.
	\end{align*}
	The zero-curvature equation $U_{t}-U_{x}+V_{x}-V_{y}+[U,V]_{s}=0$ gives
	\begin{align*}
		u_{1t_{n}}=&\sum b_{iy}\lambda^{n-i}-\sum b_{ix}\lambda^{n-i}-2\lambda\sum b_{i}\lambda^{n-i}+2u_{1}\sum a_{i}\lambda^{n-i}+2u_{4}\sum e_{i}\lambda^{n-i}+u_{1x}\\
		u_{2t_{n}}=&\sum c_{iy}\lambda^{n-i}-\sum c_{ix}\lambda^{n-i}+2\lambda\sum c_{i}\lambda^{n-i}-2u_{2}\sum a_{i}\lambda^{n-i}-2u_{3}\sum d_{i}\lambda^{n-i}+u_{2x}\\
		u_{3t_{n}}=&\sum d_{iy}\lambda^{n-i}-\sum d_{ix}\lambda^{n-i}+\lambda\sum d_{i}\lambda^{n-i}+u_{2}\sum e_{i}\lambda^{n-i}-u_{3}\sum a_{i}\lambda^{n-i}-u_{4}\sum c_{i}\lambda^{n-i}+u_{3x}\\
		u_{4t_{n}}=&\sum e_{iy}\lambda^{n-i}-\sum e_{ix}\lambda^{n-i}-\lambda\sum e_{i}\lambda^{n-i}+u_{1}\sum d_{i}\lambda^{n-i}-u_{3}\sum b_{i}\lambda^{n-i}+u_{4}\sum a_{i}\lambda^{n-i}+u_{4x}
	\end{align*}
	According to (\ref{r2}), we have
	\begin{align}
		u_{t_{n}}=\left(\begin{array}{c}
			2b_{n+1}+u_{1x}\\-2c_{n+1}+u_{2x}\\-d_{n+1}+u_{3x}\\e_{n+1}+u_{4x}
		\end{array}\right)
		=\left(\begin{array}{cccc}
			0&-2&0&0\\2&0&0&0\\0&0&0&\frac{1}{2}\\0&0&\frac{1}{2}&0
		\end{array}\right)\left(\begin{array}{c}
			-c_{n+1}\\-b_{n+1}\\2e_{n+1}\\-2d_{n+1}
		\end{array}\right)+\left(\begin{array}{c}
		u_{1x}\\u_{2x}\\u_{3x}\\u_{4x}
		\end{array}\right)=J_{2}P_{2,n+1}+u_{x}\label{h2}
	\end{align}
	where $P_{2,n+1}=(-c_{n+1},-b_{n+1},2e_{n+1},-2d_{n+1})^{t}$ and $\tilde{u}_{x}=(\frac{1}{2}u_{2x},-\frac{1}{2}u_{1x},2u_{4x},2u_{3x})^{t}$. Hierarchy (\ref{h2}) represents the $(2+1)$-dimensional generalization of the super-AKNS hierarchy.
	To construct the super-Hamiltonian structure, a direct computation yields
	\begin{align*}
		\left\langle V,\frac{\partial U}{\partial\lambda}\right\rangle=-2a,\quad \left\langle V,\frac{\partial U}{\partial u_{1}}\right\rangle=-c,\quad \left\langle V,\frac{\partial U}{\partial u_{2}}\right\rangle=-b,\quad \left\langle V,\frac{\partial U}{\partial u_{3}}\right\rangle=2e,\quad \left\langle V,\frac{\partial U}{\partial u_{4}}\right\rangle=-2d.
	\end{align*}
	Substituting the above formulate into the supertrace identity yields and balancing coefficients of each power of  in the above equality gives rise to
	\begin{align*}
		\frac{\delta}{\delta u}(-2a_{n+1})=(\tau-n)\left(\begin{array}{c}
			-c_{n}\\-b_{n}\\2e_{n}\\-2d_{n}
		\end{array}\right).
	\end{align*}
	Taking $n=1$, gives $\tau=0$. Therefore, the expression for $P_{2,n+1}$ above can be written in the following form:
	\begin{align*}
		P_{2,n+1}=\frac{2}{n+1}\frac{\delta}{\delta u}a_{n+2}
	\end{align*}
	and the hierarchy (\ref{h2}) can be expressed in the following form:
	\begin{align*}
		u_{t,n}=J_{2}P_{2,n+1}+u_{x}=J_{2}\frac{\delta H^{2}_{n+1}}{\delta u}+u_{x},\quad H^{2}_{n+1}=\frac{2}{n+1}a_{n+2}.
	\end{align*}
	This super-integrable hierarchy constitutes a (2+1)-dimemsional extension of the super-AKNS hierarchy. Evidently, its main part coincides in form with the super-AKNS hierarchy presented in Section \ref{sec2}, while an additional term involving the spatial variable is introduced.

\section{The non-isospectral superintegrable hierarchy of $\mathfrak{osp}(1,6)$ in $(2+1)$-dimension}\label{sec4}
Let $U$ and $V$ be the same as those defined in Section \ref{sec2}. consider a non-isospectral problem in $(2+1)$ dimensions
\begin{align*}
	\begin{cases}
		\phi_{\upsilon}=U\phi\\
		\phi_{\omega}=V\phi\\
		\lambda_{t}=\sum_{i\geq0}z_{i}(t)\lambda^{-i}
	\end{cases}
\end{align*}
The stationary zero curvature representation $V_{z}-V_{x}=\frac{\partial U}{\partial\lambda}\lambda_{t}+[U,V]$ gives
\begin{align}
	\begin{cases}
		a_{x}=-u_{1}q-u_{4}v-u_{5}w+u_{7}k+u_{10}n+u_{11}o+u_{13}\alpha'+u_{16}\alpha-z(t)+a_{z},\\
		b_{x}=-u_{2}r-u_{4}v-u_{6}y+u_{8}l+u_{10}n+u_{12}p+u_{14}\beta'+u_{17}\beta-z(t)+b_{z},\\
		c_{x}=-u_{3}s-u_{5}w-u_{6}y+u_{9}m+u_{11}o+u_{12}p+u_{15}\gamma'+u_{18}\gamma-z(t)+c_{z},\\
		d_{x}=-u_{1}v-u_{4}r-u_{5}y+u_{8}n+u_{10}k+u_{12}o+u_{14}\alpha'+u_{16}\beta+d_{z},\\
		e_{x}=-u_{2}v-u_{4}q-u_{6}w+u_{7}n+u_{10}l+u_{11}p+u_{13}\beta'+u_{17}\alpha+e_{z},\\
		f_{x}=-u_{1}w-u_{4}y-u_{5}s+u_{9}o+u_{11}k+u_{12}n+u_{15}\alpha'+u_{16}\gamma+f_{z},\\
		g_{x}=-u_{3}w-u_{5}q-u_{6}v+u_{7}o+u_{10}p+u_{11}m+u_{13}\gamma'+u_{18}\alpha+g_{z},\\
		h_{x}=-u_{2}y-u_{4}w-u_{6}s+u_{9}p+u_{11}n+u_{12}l+u_{15}\beta'+u_{17}\gamma+h_{z},\\
		j_{x}=-u_{3}y-u_{5}v-u_{6}r+u_{8}p+u_{10}o+u_{12}m+u_{14}\gamma'+u_{18}\beta+j_{z},\\
		k_{x}=-2\lambda k+2u_{1}a+2u_{4}d+2u_{5}f+2u_{16}\alpha'+k_{z},\\
		l_{x}=-2\lambda l+2u_{2}b+2u_{4}e+2u_{6}h+2u_{17}\beta'+l_{z},\\
		m_{x}=-2\lambda m+2u_{3}c+2u_{5}g+2u_{6}j+2u_{18}\gamma'+m_{z},\\
		n_{x}=-2\lambda n+u_{1}e+u_{2}d+u_{4}a+u_{4}b+u_{5}h+u_{6}f+u_{16}\beta'+u_{17}\alpha'+n_{z},\\
		o_{x}=-2\lambda o+u_{1}g+u_{3}f+u_{4}j+u_{5}a+u_{5}c+u_{6}d+u_{16}\gamma'+u_{18}\alpha'+o_{z},\\
		p_{x}=-2\lambda p+u_{2}j+u_{3}h+u_{4}g+u_{5}e+u_{6}b+u_{6}c+u_{17}\gamma'+u_{18}\beta'+p_{z}\\
		q_{x}=2\lambda q-2u_{7}a-2u_{10}e-2u_{11}g-2u_{13}\alpha+q_{z},\\
		r_{x}=2\lambda r-2u_{8}b-2u_{10}d-2u_{12}j-2u_{14}\beta+r_{z},\\
		s_{x}=2\lambda s-2u_{9}c-2u_{11}f-2u_{12}h-2u_{15}\gamma+s_{z},\\
		v_{x}=2\lambda v-u_{7}d-u_{8}e-u_{10}a-u_{10}b-u_{11}j-u_{12}g-u_{13}\beta-u_{14}\alpha+v_{z},\\
		w_{x}=2\lambda w-u_{7}f-u_{9}g-u_{10}h-u_{11}a-u_{11}c-u_{12}e-u_{13}\gamma-u_{15}\alpha+w_{z},\\
		y_{x}=2\lambda y-u_{8}h-u_{9}j-u_{10}f-u_{11}d-u_{12}b-u_{12}c-u_{14}\gamma-u_{15}\beta+y_{z},\\
		\alpha_{x}=\lambda\alpha+u_{7}\alpha'+u_{10}\beta'+u_{11}\gamma'-u_{13}a-u_{14}e-u_{15}g-u_{16}q-u_{17}v-u_{18}w+\alpha_{z},\\
		\beta_{x}=\lambda\beta+u_{8}\beta'+u_{10}\alpha'+u_{12}\gamma'-u_{13}d-u_{14}b-u_{15}j-u_{16}v-u_{17}r-u_{18}y+\beta_{z},\\
		\gamma_{x}=\lambda\gamma+u_{9}\gamma'+u_{11}\alpha'+u_{12}\beta'-u_{13}f-u_{14}h-u_{15}c-u_{16}w-u_{17}y-u_{18}s+\gamma_{z},\\
		\alpha'_{x}=-\lambda\alpha'+u_{1}\alpha+u_{4}\beta+u_{5}\gamma-u_{13}k-u_{14}n-u_{15}o+u_{16}a+u_{17}d+u_{18}f+\alpha'_{z},\\
		\beta'_{x}=-\lambda\beta'+u_{2}\beta+u_{4}\alpha+u_{6}\gamma-u_{13}n-u_{14}l-u_{15}p+u_{16}e+u_{17}b+u_{18}h+\beta'_{z},\\
		\gamma'_{x}=-\lambda\gamma'+u_{3}\gamma+u_{5}\alpha+u_{6}\beta-u_{13}o-u_{14}p-u_{15}m+u_{16}g+u_{17}j+u_{18}c+\gamma'_{z}.
	\end{cases}\label{r3}
\end{align}
Take the initial values
\begin{align*}
	a_{0}=b_{0}=c_{0}=1,\ d_{0}=e_{0}=\dots=w_{0}=y_{0}=\alpha_{0}=\dots=\gamma'_{0}=z_{0}(t)=0.
\end{align*}Through the recurrence relation (\ref{r3}), the leading terms of the elements in $V$ can be computed. Then, form the zero curvature equation $\frac{\partial U}{\partial u}u_{t}+\frac{\partial U}{\partial\lambda}\lambda_{t}-V_{+x}^{n}+[U,V_{+}^{n}]_{s}=0$, it follows that
\begin{align}
	u_{t,n}=\left(\begin{array}{c}
		2k_{n+1}+u_{1x}\\2l_{n+1}+u_{2x}\\2m_{n+1}+u_{3x}\\2n_{n+1}+u_{4x}\\2o_{n+1}+u_{5x}\\2p_{n+1}+u_{6x}\\-2q_{n+1}+u_{7x}\\-2r_{n+1}+u_{8x}\\-2s_{n+1}+u_{9x}\\-2v_{n+1}+u_{10x}\\-2w_{n+1}+u_{11x}\\-2y_{n+1}+u_{12x}\\-\alpha_{n+1}+u_{13x}\\-\beta_{n+1}+u_{14x}\\-\gamma_{n+1}+u_{15x}\\\alpha'_{n+1}+u_{16x}\\\beta'_{n+1}+u_{17x}\\\gamma'_{n+1}+u_{18x}\\
	\end{array}\right)=J_{1}\left(\begin{array}{c}
	-q_{n+1}\\-r_{n+1}\\-s_{n+1}\\-2v_{n+1}\\-2w_{n+1}\\-2y_{n+1}\\-k_{n+1}\\-l_{n+1}\\-m_{n+1}\\-2n_{n+1}\\-2o_{n+1}\\-2p_{n+1}\\2\alpha'_{n+1}\\2\beta'_{n+1}\\2\gamma'_{n+1}\\-2\alpha_{n+1}\\-2\beta_{n+1}\\-2\gamma_{n+1}
	\end{array}\right)+\left(\begin{array}{c}
	u_{1x}\\u_{2x}\\u_{3x}\\u_{4x}\\u_{5x}\\u_{6x}\\u_{7x}\\u_{8x}\\u_{9x}\\u_{10x}\\u_{11x}\\u_{12x}\\u_{13x}\\u_{14x}\\u_{15x}\\u_{16x}\\u_{17x}\\u_{18x}
	\end{array}\right)=J_{1}P_{3,n+1}+u_{x}.
\end{align}\label{h3}
From the recurrence relations (\ref{r3}), we have
$$	P_{3,n+1}=-LP_{3,n}-u_{L}z_{n}(t)x+P_{3,n}^{z}$$
where $P_{3,n}^{z}=(P^{z}_{1},P^{z}_{2},P^{z}_{3},P^{z}_{4},P^{z}_{5},P^{z}_{6},P^{z}_{7},P^{z}_{8},P^{z}_{9},P^{z}_{10},P^{z}_{11},P^{z}_{12},P^{z}_{13},P^{z}_{14},P^{z}_{15},P^{z}_{16},P^{z}_{17},P^{z}_{18})$ is given as follows:
\allowdisplaybreaks
\begin{align*}
		P^{z}_{1}=&-u_{7}\partial^{-1}a_{nz}-u_{10}\partial^{-1}e_{nz}-u_{11}\partial^{-1}g_{nz}-u_{13}\partial^{-1}\alpha_{nz}+\frac{1}{2}q_{nz}\\
		P^{z}_{2}=&-u_{8}\partial^{-1}b_{nz}-u_{10}\partial^{-1}d_{nz}-u_{12}\partial^{-1}j_{nz}-u_{14}\partial^{-1}\beta_{nz}+\frac{1}{2}r_{nz}\\
		P^{z}_{3}=&-u_{9}\partial^{-1}c_{nz}-u_{11}\partial^{-1}f_{nz}-u_{12}\partial^{-1}h_{nz}-u_{15}\partial^{-1}\gamma_{nz}+\frac{1}{2}s_{nz}\\
		P^{z}_{4}=&-u_{7}\partial^{-1}d_{nz}-u_{8}\partial^{-1}e_{nz}-u_{10}\partial^{-1}a_{nz}-u_{10}\partial^{-1}b_{nz}-u_{11}\partial^{-1}j_{nz}\\&-u_{12}\partial^{-1}g_{nz}-u_{13}\partial^{-1}\beta_{nz}-u_{14}\partial^{-1}\alpha_{nz}+v_{nz}\\
		P^{z}_{5}=&-u_{7}\partial^{-1}f_{nz}-u_{9}\partial^{-1}g_{nz}-u_{10}\partial^{-1}h_{nz}-u_{11}\partial^{-1}a_{nz}-u_{11}\partial^{-1}c_{nz}\\&-u_{12}\partial^{-1}e_{nz}-u_{13}\partial^{-1}\gamma_{nz}-u_{15}\partial^{-1}\alpha_{nz}+w_{nz}\\
		P^{z}_{6}=&-u_{8}\partial^{-1}h_{nz}-u_{9}\partial^{-1}j_{nz}-u_{10}\partial^{-1}f_{nz}-u_{11}\partial^{-1}d_{nz}-u_{12}\partial^{-1}b_{nz}\\&-u_{12}\partial^{-1}c_{nz}-u_{14}\partial^{-1}\gamma_{nz}-u_{15}\partial^{-1}\beta_{nz}+y_{nz}\\
		P^{z}_{7}=&-u_{1}\partial^{-1}a_{nz}-u_{4}\partial^{-1}d_{nz}-u_{5}\partial^{-1}f_{nz}-u_{16}\partial^{-1}\alpha_{nz}'-\frac{1}{2}k_{nz}\\
		P^{z}_{8}=&-u_{2}\partial^{-1}b_{nz}-u_{4}\partial^{-1}e_{nz}-u_{6}\partial^{-1}h_{nz}-u_{17}\partial^{-1}\beta_{nz}'-\frac{1}{2}l_{nz}\\
		P^{z}_{9}=&-u_{3}\partial^{-1}c_{nz}-u_{5}\partial^{-1}g_{nz}-u_{6}\partial^{-1}j_{nz}-u_{18}\partial^{-1}\gamma_{nz}'-\frac{1}{2}m_{nz}\\
		P^{z}_{10}=&-u_{1}\partial^{-1}e_{nz}-u_{2}\partial^{-1}d_{nz}-u_{4}\partial^{-1}a_{nz}-u_{4}\partial^{-1}b_{nz}-u_{5}\partial^{-1}h_{nz}\\&-u_{6}\partial^{-1}f_{nz}-u_{16}\partial^{-1}\beta_{nz}'-u_{17}\partial^{-1}\alpha_{nz}'-n_{nz}\\
		P^{z}_{11}=&-u_{1}\partial^{-1}g_{nz}-u_{3}\partial^{-1}f_{nz}-u_{4}\partial^{-1}j_{nz}-u_{5}\partial^{-1}a_{nz}-u_{5}\partial^{-1}c_{nz}\\&-u_{6}\partial^{-1}d_{nz}-u_{16}\partial^{-1}\gamma_{nz}'-u_{18}\partial^{-1}\alpha_{nz}'-o_{nz}\\
		P^{z}_{12}=&-u_{2}\partial^{-1}j_{nz}-u_{3}\partial^{-1}h_{nz}-u_{4}\partial^{-1}g_{nz}-u_{5}\partial^{-1}e_{nz}-u_{6}\partial^{-1}b_{nz}\\&-u_{6}\partial^{-1}c_{nz}-u_{17}\partial^{-1}\gamma_{nz}'-u_{18}\partial^{-1}\beta_{nz}'-p_{nz}\\
		P^{z}_{13}=&2u_{1}\partial^{-1}\alpha_{nz}+2u_{4}\partial^{-1}\beta_{nz}+2u_{5}\partial^{-1}\gamma_{nz}-2u_{13}\partial^{-1}k_{nz}-2u_{14}\partial^{-1}n_{nz}\\&-2u_{15}\partial^{-1}o_{nz}+2u_{16}\partial^{-1}a_{nz}+2u_{17}\partial^{-1}d_{nz}+2u_{18}\partial^{-1}f_{nz}+2\alpha'_{nz}\\
		P^{z}_{14}=&2u_{2}\partial^{-1}\beta_{nz}+2u_{4}\partial^{-1}\alpha_{nz}+2u_{6}\partial^{-1}\gamma_{nz}-2u_{13}\partial^{-1}n_{nz}-2u_{14}\partial^{-1}l_{nz}\\&-2u_{15}\partial^{-1}p_{nz}+2u_{16}\partial^{-1}e_{nz}+2u_{17}\partial^{-1}b_{nz}+2u_{18}\partial^{-1}h_{nz}+2\beta'_{nz}\\
		P^{z}_{15}=&2u_{3}\partial^{-1}\gamma_{nz}+2u_{5}\partial^{-1}\alpha_{nz}+2u_{6}\partial^{-1}\beta_{nz}-2u_{13}\partial^{-1}o_{nz}-2u_{14}\partial^{-1}p_{nz}\\&-2u_{15}\partial^{-1}m_{nz}+2u_{16}\partial^{-1}g_{nz}+2u_{17}\partial^{-1}j_{nz}+2u_{18}\partial^{-1}c_{nz}+2\gamma'_{nz}\\
		P^{z}_{16}=&2u_{7}\partial^{-1}\alpha_{nz}'+2u_{10}\partial^{-1}\beta_{nz}'+2u_{11}\partial^{-1}\gamma_{nz}'-2u_{13}\partial^{-1}a_{nz}-2u_{14}\partial^{-1}e_{nz}\\&-2u_{15}\partial^{-1}g_{nz}-2u_{16}\partial^{-1}q_{nz}-2u_{17}\partial^{-1}v_{nz}-2u_{18}\partial^{-1}w_{nz}+2\alpha_{nz}\\
		P^{z}_{17}=&2u_{8}\partial^{-1}\beta_{nz}'+2u_{10}\partial^{-1}\alpha_{nz}'+2u_{12}\partial^{-1}\gamma_{nz}'-2u_{13}\partial^{-1}d_{nz}-2u_{14}\partial^{-1}b_{nz}\\&-2u_{15}\partial^{-1}j_{nz}-2u_{16}\partial^{-1}v_{nz}-2u_{17}\partial^{-1}r_{nz}-2u_{18}\partial^{-1}y_{nz}+2\beta_{nz}\\
		P^{z}_{18}=&2u_{9}\partial^{-1}\gamma_{nz}'+2u_{11}\partial^{-1}\alpha_{nz}'+2u_{12}\partial^{-1}\beta_{nz}'-2u_{13}\partial^{-1}f_{nz}-2u_{14}\partial^{-1}h_{nz}\\&-2u_{15}\partial^{-1}c_{nz}-2u_{16}\partial^{-1}w_{nz}-2u_{17}\partial^{-1}y_{nz}-2u_{18}\partial^{-1}s_{nz}+2\gamma_{nz}
\end{align*}
Based on the super‑trace identity and the calculations presented in above sections, we obtain
\begin{align*}
	&\frac{\delta}{\delta u}(14a_{n+1}14b_{n+1}14c_{n+1})\\&=(\tau-n)
		(7q_{n},\  7r_{n},\  7s_{n},\  14v_{n},\  14w_{n},\  14y_{n},\  7k_{n},\  7l_{n},\  7m_{n},\  
		14n_{n},\  14o_{n},\  14p_{n},\  -14\alpha'_{n},\  -14\beta'_{n},\  -14\gamma'_{n},\  14\alpha_{n},\  14\beta_{n},\  14\gamma_{n}).
\end{align*}
Taking $n=1$ and using the above equation along with the leading terms of $a,\cdots,\gamma$, we obtain $\tau=0$. Therefore, the expression for $P_{3,n+1}$ above can be written in the following form:
\begin{align*}
	P_{3,n+1}=\frac{\delta}{\delta u}(\frac{2}{n+1}(a_{n+2}+b_{n+2}+c_{n+2})).
\end{align*}
Therefore, the integrable hierarchy (\ref{h3}) can be expressed as
\begin{align*}
	u_{t,n}=J_{1}P_{3,n+1}+u_{x}=J_{1}\frac{\delta H^{3}_{n+1}}{\delta u}+u_{x},\ H^{3}_{n+1}=\frac{2}{n+1}(a_{n+2}+b_{n+2}+c_{n+2})
\end{align*}
By employing a $(2+1)$-dimensional non‑isospectral problem, we have derived a family of $(2+1)$-dimensional non‑isospectral integrable hierarchies on osp.

\section{Concluding Remarks}
Lie superalgebras originate from the mathematical abstraction of graded structures and the need for describing supersymmetry in physics, serving as a classic example of the intersection between mathematics and physics. In theoretical physics, there exists a foundational principle known as Noether's theorem, which establishes a one‑to‑one correspondence between symmetries and conservation laws (conserved quantities, constants of motion). Conserved quantities (constants of motion), in turn, are inherently linked to integrable systems. To explore the connection between supersymmetry and super-integrability, we choose to construct super-integrable systems starting from specific Lie superalgebras. Prior to this, scholars have constructed several integrable systems on Lie superalgebras such as $B(0,1)$, $\mathfrak{sl}(2,\mathbb{R})$, and $\mathfrak{sl}(2N,1)$. Therefore, we choose to construct a new super-integrable system and its super-Hamilton structure on $\mathfrak{osp}(l, r)$, which belongs to one of the four infinite families of Lie superalgebras. Upon completing the above work, we have indeed constructed a new super‑integrable system and identified certain connections between it and the super‑AKNS hierarchy. However, focusing on super‑integrable systems derived from a specific Lie superalgebra imposes inherent limitations. In our subsequent research, we intend to shift toward uncovering the general relationship between the supersymmetry of Lie superalgebras and super‑integrability, rather than confining the study to any particular Lie superalgebra.

\section*{DECLARATIONS}
\subsection*{Ethics approval and consent to participate}\vspace{-1.5em}
This study did not involve human participants, animal subjects, or any tissue samples, and therefore did not require ethical approval.\vspace{-1.5em}
\subsection*{Consent for publication}\vspace{-1.5em}
Consent for publication was obtained from all individual participants included in the study.\vspace{-1.5em}
\subsection*{Availability of data and materials}\vspace{-1.5em}
Not applicable.\vspace{-1.5em}
\subsection*{Competing interests}\vspace{-1.5em}
The authors declare that they have no competing interests.\vspace{-1.5em}
\subsection*{Funding }\vspace{-1.5em}
This research was supported by the National Natural Science Foundation of China (No. 11961049, 10601219) and by the Key Project of Jiangxi Natural Science Foundation grant (No. 20232ACB201004). The funding body played no role in the design of the study and collection, analysis, and interpretation of data and in writing the manuscript.\vspace{-1.5em}

\begin{acknowledgments}
This research was supported by the National Natural Science Foundation of China (No. 11961049, 10601219) and by the Key Project of Jiangxi Natural Science Foundation grant (No. 20232ACB201004).
\end{acknowledgments}

\appendix
\section{\label{appA}The remaining structure constants of $\mathfrak{osp}(1,6)$}
\begin{align*}
	&[E_{1}, E_{4}]=E_{4},\
	[E_{1}, E_{5}]=-E_{5},\
	[E_{1}, E_{6}]=E_{6},\
	[E_{1}, E_{7}]=-E_{7},\
	[E_{1}, E_{10}]=2E_{10},\
	[E_{1}, E_{13}]=E_{13},\
	[E_{1}, E_{14}]=E_{14},\\&
	[E_{1}, E_{16}]=-2E_{16},\
	[E_{1}, E_{19}]=-E_{19},\
	[E_{1}, E_{20}]=-E_{20},\
	[E_{1}, E_{22}]=-E_{22},\
	[E_{1}, E_{25}]=E_{25},\
	[E_{2}, E_{4}]=-E_{4},\\&
	[E_{2}, E_{5}]=E_{5},\
	[E_{2}, E_{8}]=E_{8},\
	[E_{2}, E_{9}]=-E_{9},\
	[E_{2}, E_{11}]=2E_{11},\
	[E_{2}, E_{13}]=E_{13},\
	[E_{2}, E_{15}]=E_{15},\
	[E_{2}, E_{17}]=-2E_{17},\\&
	[E_{2}, E_{19}]=-E_{19},\
	[E_{2}, E_{21}]=-E_{21},\
	[E_{2}, E_{23}]=-E_{23},\
	[E_{2}, E_{26}]=E_{26},\
	[E_{3}, E_{6}]=-E_{6},\
	[E_{3}, E_{7}]=E_{7},\
	[E_{3}, E_{8}]=-E_{8},\\&
	[E_{3}, E_{9}]=E_{9},\
	[E_{3}, E_{12}]=2E_{12},\
	[E_{3}, E_{14}]=E_{14},\
	[E_{3}, E_{15}]=E_{15},\
	[E_{3}, E_{18}]=-2E_{18},\
	[E_{3}, E_{20}]=-E_{20},\
	[E_{3}, E_{21}]=-E_{21},\\&
	[E_{3}, E_{24}]=-E_{24},\
	[E_{3}, E_{27}]=E_{27},\
	[E_{4}, E_{5}]=E_{1}-E_{2},\
	[E_{4}, E_{7}]=-E_{9},\
	[E_{4}, E_{8}]=E_{6},\
	[E_{4}, E_{11}]=E_{13},\
	[E_{4}, E_{13}]=2E_{10},\\&
	[E_{4}, E_{15}]=E_{14},\
	[E_{4}, E_{16}]=-E_{19},\
	[E_{4}, E_{19}]=-2E_{17},\
	[E_{4}, E_{20}]=-E_{21},\
	[E_{4}, E_{22}]=-E_{23},\
	[E_{4}, E_{26}]=E_{25},\\&
	[E_{5}, E_{6}]=E_{8},\
	[E_{5}, E_{9}]=-E_{7},\
	[E_{5}, E_{10}]=E_{13},\
	[E_{5}, E_{13}]=2E_{11},\
	[E_{5}, E_{14}]=E_{15},\
	[E_{5}, E_{17}]=-E_{19},\
	[E_{5}, E_{19}]=-2E_{16},\\&
	[E_{5}, E_{21}]=-E_{20},\
	[E_{5}, E_{23}]=-E_{22},\
	[E_{5}, E_{25}]=E_{26},\
	[E_{6}, E_{7}]=E_{1}-E_{3},\
	[E_{6}, E_{9}]=E_{4},\
	[E_{6}, E_{12}]=E_{14},\
	[E_{6}, E_{14}]=2E_{10},\\&
	[E_{6}, E_{15}]=E_{13},\
	[E_{6}, E_{16}]=-E_{20},\
	[E_{6}, E_{19}]=-E_{21},\
	[E_{6}, E_{20}]=-2E_{18},\
	[E_{6}, E_{22}]=-E_{24},\
	[E_{6}, E_{27}]=E_{25},\
	[E_{7}, E_{8}]=-E_{5},\\&
	[E_{7}, E_{10}]=E_{14},\
	[E_{7}, E_{13}]=E_{15},\
	[E_{7}, E_{14}]=2E_{12},\
	[E_{7}, E_{18}]=-E_{20},\
	[E_{7}, E_{20}]=-2E_{16},\
	[E_{7}, E_{21}]=-E_{19},\
	[E_{7}, E_{24}]=-E_{22},\\&
	[E_{7}, E_{25}]=E_{27},\
	[E_{8}, E_{9}]=E_{2}-E_{3},\
	[E_{8}, E_{12}]=E_{15},\
	[E_{8}, E_{14}]=E_{13},\
	[E_{8}, E_{15}]=2E_{11},\
	[E_{8}, E_{17}]=-E_{21},\
	[E_{8}, E_{19}]=-E_{20},\\&
	[E_{8}, E_{21}]=-2E_{18},\
	[E_{8}, E_{23}]=-E_{24},\
	[E_{8}, E_{27}]=E_{26},\
	[E_{9}, E_{11}]=E_{15},\
	[E_{9}, E_{13}]=E_{14},\
	[E_{9}, E_{15}]=2E_{12},\
	[E_{9}, E_{18}]=-E_{21},\\&
	[E_{9}, E_{20}]=-E_{19},\
	[E_{9}, E_{21}]=-2E_{17},\
	[E_{9}, E_{24}]=-E_{23},\
	[E_{9}, E_{26}]=E_{27},\
	[E_{10}, E_{16}]=E_{1},\
	[E_{10}, E_{19}]=E_{4},\
	[E_{10}, E_{20}]=E_{6},\\&
	[E_{10}, E_{22}]=-E_{25},\
	[E_{11}, E_{17}]=E_{2},\
	[E_{11}, E_{19}]=E_{5},\
	[E_{11}, E_{21}]=E_{8},\
	[E_{11}, E_{23}]=-E_{26},\
	[E_{12}, E_{18}]=E_{3},\
	[E_{12}, E_{20}]=E_{7},\\&
	[E_{12}, E_{21}]=E_{9},\
	[E_{12}, E_{24}]=-E_{27},\
	[E_{13}, E_{16}]=E_{5},\
	[E_{13}, E_{17}]=E_{4},\
	[E_{13}, E_{19}]=E_{1}+E_{2},\
	[E_{13}, E_{20}]=E_{8},\
	[E_{13}, E_{21}]=E_{6},\\&
	[E_{13}, E_{22}]=-E_{26},\
	[E_{13}, E_{23}]=-E_{25},\
	[E_{14}, E_{16}]=E_{7},\
	[E_{14}, E_{18}]=E_{6},\
	[E_{14}, E_{19}]=E_{9},\
	[E_{14}, E_{20}]=E_{1}+E_{3},\
	[E_{14}, E_{21}]=E_{4},\\&
	[E_{14}, E_{22}]=-E_{27},\
	[E_{14}, E_{24}]=-E_{25},\
	[E_{15}, E_{17}]=E_{9},\
	[E_{15}, E_{18}]=E_{8},\
	[E_{15}, E_{19}]=E_{7},\
	[E_{15}, E_{20}]=E_{5},\
	[E_{15}, E_{21}]=E_{2}+E_{3},\\&
	[E_{15}, E_{23}]=-E_{27},\
	[E_{15}, E_{24}]=-E_{26},\
	[E_{16}, E_{25}]=-E_{22},\
	[E_{17}, E_{26}]=-E_{23},\
	[E_{18}, E_{27}]=-E_{24},\
	[E_{19}, E_{25}]=-E_{23},\\&
	[E_{19}, E_{26}]=-E_{22},\
	[E_{20}, E_{25}]=-E_{24},\
	[E_{20}, E_{27}]=-E_{22},\
	[E_{21}, E_{26}]=-E_{24},\
	[E_{21}, E_{27}]=-E_{23},\
	[E_{22}, E_{22}]_{+}=2E_{16},\\&
	[E_{22}, E_{23}]_{+}=E_{19},\
	[E_{22}, E_{24}]_{+}=E_{20},\
	[E_{22}, E_{25}]_{+}=-E_{1},\
	[E_{22}, E_{26}]_{+}=-E_{5},\
	[E_{22}, E_{27}]_{+}=-E_{7},\
	[E_{23}, E_{23}]_{+}=2E_{17},\\&
	[E_{23}, E_{24}]_{+}=E_{21},\
	[E_{23}, E_{25}]_{+}=-E_{4},\
	[E_{23}, E_{26}]_{+}=-E_{2},\
	[E_{23}, E_{27}]_{+}=-E_{9},\
	[E_{24}, E_{24}]_{+}=2E_{18},\
	[E_{24}, E_{25}]_{+}=-E_{6},\\&
	[E_{24}, E_{26}]_{+}=-E_{8},\
	[E_{24}, E_{27}]_{+}=-E_{3},\
	[E_{25}, E_{25}]_{+}=-2E_{10},\
	[E_{25}, E_{26}]_{+}=-E_{13},\
	[E_{25}, E_{27}]_{+}=-E_{14},\
	[E_{26}, E_{26}]_{+}=-2E_{11},\\&
	[E_{26}, E_{27}]_{+}=-E_{15},\
	[E_{27}, E_{27}]_{+}=-2E_{12},\
\end{align*}

\section{\label{appB}Recurrence Operator $P_{1,n+1}=L_{1}P_{1,n}+u_{L}z_{n}(t)x$}
\allowdisplaybreaks
\begin{align*}
	&l_{1,1}=-\frac{1}{2}\partial+u_{7}\partial^{-1}u_{1}+u_{10}\partial^{-1}u_{4}+u_{11}\partial^{-1}u_{5},\ l_{1,4}=\frac{1}{2}u_{7}\partial^{-1}u_{4}+\frac{1}{2}u_{10}\partial^{-1}u_{2}+\frac{1}{2}u_{11}\partial^{-1}u_{6},\\
	&l_{1,5}=\frac{1}{2}u_{7}\partial^{-1}u_{5}+\frac{1}{2}u_{10}\partial^{-1}u_{6}+\frac{1}{2}u_{11}\partial^{-1}u_{3},\ l_{1,7}=-u_{7}\partial^{-1}u_{7},\ l_{1,8}=-u_{10}\partial^{-1}u_{10},\ l_{1,9}u_{11}\partial^{-1}u_{11},\\& l_{1,10}=-\frac{1}{2}u_{7}\partial^{-1}u_{10}-\frac{1}{2}u_{10}\partial^{-1}u_{7},\ l_{1,11}=-\frac{1}{2}u_{7}\partial^{-1}u_{11}-\frac{1}{2}u_{11}\partial^{-1}u_{7},\ l_{1,12}=-\frac{1}{2}u_{10}\partial^{-1}u_{11}-\frac{1}{2}u_{11}\partial^{-1}u_{10},\\& l_{1,13}=\frac{1}{2}u_{7}\partial^{-1}u_{13},\ l_{1,14}=\frac{1}{2}u_{10}\partial^{-1}u_{13},\ l_{1,15}=\frac{1}{2}u_{11}\partial^{-1}u_{13},\ l_{1,16}=\frac{1}{2}u_{13}-\frac{1}{2}u_{7}\partial^{-1}u_{16}-\frac{1}{2}u_{10}\partial^{-1}u_{17}-\frac{1}{2}u_{11}\partial^{-1}u_{18},\\&
	l_{2,2}=-\frac{1}{2}\partial+u_{8}\partial^{-1}u_{2}+u_{10}\partial^{-1}u_{4}+u_{12}\partial^{-1}u_{6},\ l_{2,4}=\frac{1}{2}u_{8}\partial^{-1}u_{4}+\frac{1}{2}u_{10}\partial^{-1}u_{1}+\frac{1}{2}u_{12}\partial^{-1}u_{5},\\& l_{2,6}=\frac{1}{2}u_{8}\partial^{-1}u_{6}+\frac{1}{2}u_{10}\partial^{-1}u_{5}+\frac{1}{2}u_{12}\partial^{-1}u_{3},\ l_{2,7}=-u_{10}\partial^{-1}u_{10},\ l_{2,8}=-u_{8}\partial^{-1}u_{8},\ l_{2,9}=-u_{12}\partial^{-1}u_{12},\\& l_{2,10}=-\frac{1}{2}u_{8}\partial^{-1}u_{10}-\frac{1}{2}u_{10}\partial^{-1}u_{8},\ l_{2,11}=-\frac{1}{2}u_{10}\partial^{-1}u_{12}-\frac{1}{2}u_{12}\partial^{-1}u_{10},\ l_{2,12}=-\frac{1}{2}u_{8}\partial^{-1}u_{12}-\frac{1}{2}u_{12}\partial^{-1}u_{8},\\& l_{2,13}=\frac{1}{2}u_{10}\partial^{-1}u_{14},\ l_{2,14}=\frac{1}{2}u_{8}\partial^{-1}u_{14},\ l_{2,15}=\frac{1}{2}u_{12}\partial^{-1}u_{14},\ l_{2,17}=\frac{1}{2}u_{14}-\frac{1}{2}u_{8}\partial^{-1}u_{17}-\frac{1}{2}u_{10}\partial^{-1}u_{16}-\frac{1}{2}u_{12}\partial^{-1}u_{18},\\&
	l_{3,3}=-\frac{1}{2}\partial+u_{9}\partial^{-1}u_{3}+u_{11}\partial^{-1}u_{5}+u_{12}\partial^{-1}u_{6},\ l_{3,5}=\frac{1}{2}u_{9}\partial^{-1}u_{5}+\frac{1}{2}u_{11}\partial^{-1}u_{1}+\frac{1}{2}u_{12}\partial^{-1}u_{4},\\& l_{3,6}=\frac{1}{2}u_{9}\partial^{-1}u_{6}+\frac{1}{2}u_{11}\partial^{-1}u_{4}+\frac{1}{2}u_{12}\partial^{-1}u_{2},\ l_{3,7}=-u_{11}\partial^{-1}u_{11},\ l_{3,8}=-u_{12}\partial^{-1}u_{12},\ l_{3,9}=-u_{9}\partial^{-1}u_{9},\\& l_{3,10}=-\frac{1}{2}u_{11}\partial^{-1}u_{12}-\frac{1}{2}u_{12}\partial^{-1}u_{11},\ l_{3,11}=-\frac{1}{2}u_{9}\partial^{-1}u_{11}-\frac{1}{2}u_{11}\partial^{-1}u_{9},\ l_{3,12}=-\frac{1}{2}u_{9}\partial^{-1}u_{12}-\frac{1}{2}u_{12}\partial^{-1}u_{9},\\& l_{3,13}=-\frac{1}{2}u_{11}\partial^{-1}u_{15},\ l_{3,14}=-\frac{1}{2}u_{12}\partial^{-1}u_{15},\ l_{3,15}=-\frac{1}{2}u_{9}\partial^{-1}u_{15},\ l_{3,18}=\frac{1}{2}u_{15}-\frac{1}{2}u_{9}\partial^{-1}u_{18}-\frac{1}{2}u_{11}\partial^{-1}u_{16}-\frac{1}{2}u_{12}\partial^{-1}u_{17},\\&
	l_{4,1}=u_{8}\partial^{-1}u_{4}+u_{10}\partial^{-1}u_{1}+u_{12}\partial^{-1}u_{5},\ l_{4,2}=u_{7}\partial^{-1}u_{4}+u_{10}\partial^{-1}u_{2}+u_{11}\partial^{-1}u_{6},\\& l_{4,4}=-\partial+\frac{1}{2}u_{7}\partial^{-1}u_{1}+\frac{1}{2}u_{8}\partial^{-1}u_{2}+u_{10}\partial^{-1}u_{4}+\frac{1}{2}u_{11}\partial^{-1}u_{5}+\frac{1}{2}u_{12}\partial^{-1}u_{6},\ l_{4,5}=\frac{1}{2}u_{8}\partial^{-1}u_{6}+\frac{1}{2}u_{10}\partial^{-1}u_{5}+\frac{1}{2}u_{12}\partial^{-1}u_{3},\\& l_{4,6}=\frac{1}{2}u_{7}\partial^{-1}u_{5}+\frac{1}{2}u_{10}\partial^{-1}u_{6}+\frac{1}{2}u_{11}\partial^{-1}u_{3},\ l_[4,7]=-u_{7}\partial^{-1}u_{10}-u_{10}\partial^{-1}u_{7},\ l_{4,8}=-u_{8}\partial^{-1}u_{10}-u_{10}\partial^{-1}u_{8},\\& l_{4,9}=-u_{11}\partial^{-1}u_{12}-u_{12}\partial^{-1}u_{11},\ l_{4,10}=-\frac{1}{2}u_{7}\partial^{-1}u_{8}-\frac{1}{2}u_{8}\partial^{-1}u_{7}-u_{10}\partial^{-1}u_{10},\\& l_{4,11}=-\frac{1}{2}u_{7}\partial^{-1}u_{12}-\frac{1}{2}u_{10}\partial^{-1}u_{11}-\frac{1}{2}u_{11}\partial^{-1}u_{10}-\frac{1}{2}u_{12}\partial^{-1}u_{7},\\& l_{4,12}=-\frac{1}{2}u_{8}\partial^{-1}u_{11}-\frac{1}{2}u_{10}\partial^{-1}u_{12}-\frac{1}{2}u_{11}\partial^{-1}u_{8}-\frac{1}{2}u_{12}\partial^{-1}u_{10},\\& l_{4,13}=\frac{1}{2}u_{7}\partial^{-1}u_{14}+\frac{1}{2}u_{10}\partial^{-1}u_{13},\ l_{4,14}=\frac{1}{2}u_{8}\partial^{-1}u_{13}+\frac{1}{2}u_{10}\partial^{-1}u_{14},\ l_{4,15}=\frac{1}{2}u_{11}\partial^{-1}u_{14}+\frac{1}{2}u_{12}\partial^{-1}u_{13},\\& l_{4,16}=\frac{1}{2}u_{14}-\frac{1}{2}u_{8}\partial^{-1}u_{17}-\frac{1}{2}u_{10}\partial^{-1}u_{16}-\frac{1}{2}u_{12}\partial^{-1}u_{18},\ l_{4,17}=\frac{1}{2}u_{13}-\frac{1}{2}u_{7}\partial^{-1}u_{16}-\frac{1}{2}u_{10}\partial^{-1}u_{17}-\frac{1}{2}u_{11}\partial^{-1}u_{18},\\&
	l_{5,1}=u_{9}\partial^{-1}u_{5}+u_{11}\partial^{-1}u_{1}+u_{12}\partial^{-1}u_{4},\ l_{5,3}=u_{7}\partial^{-1}u_{5}+u_{10}\partial^{-1}u_{6}+u_{11}\partial^{-1}u_{3},\\& l_{5,4}=\frac{1}{2}u_{9}\partial^{-1}u_{6}+\frac{1}{2}u_{11}\partial^{-1}u_{4}+\frac{1}{2}u_{12}\partial^{-1}u_{2},\ l_{5,5}=-\frac{1}{2}\partial+\frac{1}{2}u_{7}\partial^{-1}u_{1}+\frac{1}{2}u_{9}\partial^{-1}u_{3}+\frac{1}{2}u_{10}\partial^{-1}u_{4}+u_{11}\partial^{-1}u_{5}+\frac{1}{2}u_{12}\partial^{-1}u_{6},\\& l_{5,6}=\frac{1}{2}u_{7}\partial^{-1}u_{4}+\frac{1}{2}u_{10}\partial^{-1}u_{2}+\frac{1}{2}u_{11}\partial^{-1}u_{6},\ l_{5,7}=-u_{7}\partial^{-1}u_{11}-u_{11}\partial^{-1}u_{7},\ l_{5,8}=-u_{10}\partial^{-1}u_{12}-u_{12}\partial^{-1}u_{10},\\& l_{5,9}=-u_{9}\partial^{-1}u_{11}-u_{11}\partial^{-1}u_{9},\ l_{5,10}=-\frac{1}{2}u_{7}\partial^{-1}u_{12}-\frac{1}{2}u_{10}\partial^{-1}u_{11}-\frac{1}{2}u_{11}\partial^{-1}u_{10}-\frac{1}{2}u_{12}\partial^{-1}u_{7},\\& l_{5,11}=-\frac{1}{2}u_{7}\partial^{-1}u_{9}-\frac{1}{2}u_{9}\partial^{-1}u_{7}-u_{11}\partial^{-1}u_{11},\ l_{5,12}=-\frac{1}{2}u_{9}\partial^{-1}u_{10}-\frac{1}{2}u_{10}\partial^{-1}u_{9}-\frac{1}{2}u_{11}\partial^{-1}u_{12}-\frac{1}{2}u_{12}\partial^{-1}u_{11},\\& l_{5,13}=\frac{1}{2}u_{7}\partial^{-1}u_{15}+\frac{1}{2}u_{11}\partial^{-1}u_{13},\ l_{5,14}=\frac{1}{2}u_{10}\partial^{-1}u_{15}+\frac{1}{2}u_{12}\partial^{-1}u_{13},\ l_{5,15}=\frac{1}{2}u_{9}\partial^{-1}u_{13}+\frac{1}{2}u_{11}\partial^{-1}u_{15},\\& l_{5,16}=\frac{1}{2}u_{15}-\frac{1}{2}u_{9}\partial^{-1}u_{18}-\frac{1}{2}u_{11}\partial^{-1}u_{16}-\frac{1}{2}u_{12}\partial^{-1}u_{17},\ l_{5,18}=\frac{1}{2}u_{13}-\frac{1}{2}u_{7}\partial^{-1}u_{16}-\frac{1}{2}u_{10}\partial^{-1}u_{17}-\frac{1}{2}u_{11}\partial^{-1}u_{18},\\&
	l_{6,2}=u_{9}\partial^{-1}u_{6}+u_{11}\partial^{-1}u_{4}+u_{12}\partial^{-1}u_{2},\ l_{6,3}=u_{8}\partial^{-1}u_{6}+u_{10}\partial^{-1}u_{5}+u_{12}\partial^{-1}u_{3},\ l_{6,4}=\frac{1}{2}u_{9}\partial^{-1}u_{5}+\frac{1}{2}u_{11}\partial^{-1}u_{1}+\frac{1}{2}u_{12}\partial^{-1}u_{4},\\& l_{6,5}=\frac{1}{2}u_{8}\partial^{-1}u_{4}+\frac{1}{2}u_{10}\partial^{-1}u_{1}+\frac{1}{2}u_{12}\partial^{-1}u_{5},\ l_{6,6}=-\frac{1}{2}\partial+\frac{1}{2}u_{8}\partial^{-1}u_{2}+\frac{1}{2}u_{9}\partial^{-1}u_{3}+\frac{1}{2}u_{10}\partial^{-1}u_{4}+\frac{1}{2}u_{11}\partial^{-1}u_{5}+u_{12}\partial^{-1}u_{6},\\& l_{6,7}=-u_{10}\partial^{-1}u_{11}-u_{11}\partial^{-1}u_{10},\ l_{6,8}=-u_{8}\partial^{-1}u_{12}-u_{12}\partial^{-1}u_{8},\ l_{6,9}=-u_{9}\partial^{-1}u_{12}-u_{12}\partial^{-1}u_{9},\\& l_{6,10}=-\frac{1}{2}u_{8}\partial^{-1}u_{11}-\frac{1}{2}u_{10}\partial^{-1}u_{12}-\frac{1}{2}u_{11}\partial^{-1}u_{8}-\frac{1}{2}u_{12}\partial^{-1}u_{10},\\& l_{6,11}=-\frac{1}{2}u_{9}\partial^{-1}u_{10}-\frac{1}{2}u_{10}\partial^{-1}u_{9}-\frac{1}{2}u_{11}\partial^{-1}u_{12}-\frac{1}{2}u_{12}\partial^{-1}u_{11},\\& l_{6,12}=-\frac{1}{2}u_{8}\partial^{-1}u_{9}-\frac{1}{2}u_{9}\partial^{-1}u_{8}-u_{12}\partial^{-1}u_{12},\ l_{6,13}=\frac{1}{2}u_{10}\partial^{-1}u_{15}+\frac{1}{2}u_{11}\partial^{-1}u_{14},\ l_{6,14}=\frac{1}{2}u_{8}\partial^{-1}u_{15}+\frac{1}{2}u_{12}\partial^{-1}u_{14},\\& l_{6,15}=\frac{1}{2}u_{9}\partial^{-1}u_{14}+\frac{1}{2}u_{12}\partial^{-1}u_{15},\ l_{6,17}=\frac{1}{2}u_{15}-\frac{1}{2}u_{9}\partial^{-1}u_{18}-\frac{1}{2}u_{11}\partial^{-1}u_{16}-\frac{1}{2}u_{12}\partial^{-1}u_{17},\\& l_{6,18}=\frac{1}{2}u_{14}-\frac{1}{2}u_{8}\partial^{-1}u_{17}-\frac{1}{2}u_{10}\partial^{-1}u_{16}-\frac{1}{2}u_{12}\partial^{-1}u_{18},\\& 
	l_{7,1}=u_{1}\partial^{-1}u_{1},\ l_{7,2}=u_{4}\partial^{-1}u_{4},\ l_{7,3}=u_{5}\partial^{-1}u_{5},\ l_{7,4}=\frac{1}{2}u_{1}\partial^{-1}u_{4}+\frac{1}{2}u_{4}\partial^{-1}u_{1},\ l_{7,5}=\frac{1}{2}u_{1}\partial^{-1}u_{5}+\frac{1}{2}u_{5}\partial^{-1}u_{1},\\& l_{7,6}=\frac{1}{2}u_{4}\partial^{-1}u_{5}+\frac{1}{2}u_{5}\partial^{-1}u_{4},\ l_{7,7}=\frac{1}{2}\partial-u_{1}\partial^{-1}u_{7}-u_{4}\partial^{-1}u_{10}-u_{5}\partial^{-1}u_{11},\\& l_{7,10}=-\frac{1}{2}u_{1}\partial^{-1}u_{10}-\frac{1}{2}u_{4}\partial^{-1}u_{8}-\frac{1}{2}u_{5}\partial^{-1}u_{12},\ l_{7,11}=-\frac{1}{2}u_{1}\partial^{-1}u_{11}-\frac{1}{2}u_{4}\partial^{-1}u_{12}-\frac{1}{2}u_{5}\partial^{-1}u_{9},\\& l_{7,13}=-\frac{1}{2}u_{16}+\frac{1}{2}u_{1}\partial^{-1}u_{13}+\frac{1}{2}u_{4}\partial^{-1}u_{14}+\frac{1}{2}u_{5}\partial^{-1}u_{15},\ l_{7,16}=-\frac{1}{2}u_{1}\partial^{-1}u_{16},\\& l_{7,17}=-\frac{1}{2}u_{4}\partial^{-1}u_{16},\ l_{7,18}=-\frac{1}{2}u_{5}\partial^{-1}u_{16},\ 
	l_{8,1}=u_{4}\partial^{-1}u_{4},\ l_{8,2}=u_{2}\partial^{-1}u_{2},\ l_{8,3}=u_{6}\partial^{-1}u_{6},\ l_{8,4}=\frac{1}{2}u_{2}\partial^{-1}u_{4}+\frac{1}{2}u_{4}\partial^{-1}u_{2},\\& l_{8,5}=\frac{1}{2}u_{4}\partial^{-1}u_{6}+\frac{1}{2}u_{6}\partial^{-1}u_{4},\ l_{8,6}=\frac{1}{2}u_{2}\partial^{-1}u_{6}+\frac{1}{2}u_{6}\partial^{-1}u_{2},\ l_{8,8}=\frac{1}{2}\partial-u_{2}\partial^{-1}u_{8}-u_{4}\partial^{-1}u_{10}-u_{6}\partial^{-1}u_{12},\\& l_{8,10}=-\frac{1}{2}u_{2}\partial^{-1}u_{10}-\frac{1}{2}u_{4}\partial^{-1}u_{7}-\frac{1}{2}u_{6}\partial^{-1}u_{11},\ l_{8,12}=-\frac{1}{2}u_{2}\partial^{-1}u_{12}-\frac{1}{2}u_{4}\partial^{-1}u_{11}-\frac{1}{2}u_{6}\partial^{-1}u_{9},\\& l_{8,14}=-\frac{1}{2}u_{17}+\frac{1}{2}u_{2}\partial^{-1}u_{14}+\frac{1}{2}u_{4}\partial^{-1}u_{13}+\frac{1}{2}u_{6}\partial^{-1}u_{15},\ l_{8,15}=-\frac{1}{2}u_{4}\partial^{-1}u_{17},\\& l_{8,17}=-\frac{1}{2}u_{2}\partial^{-1}u_{17},\ l_{8,18}=-\frac{1}{2}u_{6}\partial^{-1}u_{17},\ 
	l_{9,1}=u_{5}\partial^{-1}u_{5},\ l_{9,2}=u_{6}\partial^{-1}u_{6},\ l_{9,3}=u_{3}\partial^{-1}u_{3},\\& l_{9,4}=\frac{1}{2}u_{5}\partial^{-1}u_{6}+\frac{1}{2}u_{6}\partial^{-1}u_{5},\ l_{9,5}=\frac{1}{2}u_{3}\partial^{-1}u_{5}+\frac{1}{2}u_{5}\partial^{-1}u_{3},\\& l_{9,6}=\frac{1}{2}u_{3}\partial^{-1}u_{6}+\frac{1}{2}u_{6}\partial^{-1}u_{3},\ l_{9,9}=\frac{1}{2}\partial-u_{3}\partial^{-1}u_{9}-u_{5}\partial^{-1}u_{11}-u_{6}\partial^{-1}u_{12},\ l_{9,11}=-\frac{1}{2}u_{3}\partial^{-1}u_{11}-\frac{1}{2}u_{5}\partial^{-1}u_{7}-\frac{1}{2}u_{6}\partial^{-1}u_{10},\\& l_{9,12}=-\frac{1}{2}u_{3}\partial^{-1}u_{12}-\frac{1}{2}u_{5}\partial^{-1}u_{10}-\frac{1}{2}u_{6}\partial^{-1}u_{8},\ l_{9,15}=-\frac{1}{2}u_{18}+\frac{1}{2}u_{3}\partial^{-1}u_{15}+\frac{1}{2}u_{5}\partial^{-1}u_{13}+\frac{1}{2}u_{6}\partial^{-1}u_{14},\\& l_{9,16}=-\frac{1}{2}u_{5}\partial^{-1}u_{18},\ l_{9,17}=-\frac{1}{2}u_{6}\partial^{-1}u_{18},\ l_{9,18}=-\frac{1}{2}u_{3}\partial^{-1}u_{12},\\&
	l_{10,1}=u_{1}\partial^{-1}u_{4}+u_{4}\partial^{-1}u_{1},\ l_{10,2}=u_{2}\partial^{-1}u_{4}+u_{4}\partial^{-1}u_{2},\ l_{10,3}=u_{5}\partial^{-1}u_{6}+u_{6}\partial^{-1}u_{5},\\& l_{10,4}=\frac{1}{2}u_{1}\partial^{-1}u_{2}+\frac{1}{2}u_{2}\partial^{-1}u_{1}+u_{4}\partial^{-1}u_{4},\ l_{10,5}=\frac{1}{2}u_{1}\partial^{-1}u_{6}+\frac{1}{2}u_{4}\partial^{-1}u_{5}+\frac{1}{2}u_{5}\partial^{-1}u_{4}+\frac{1}{2}u_{6}\partial^{-1}u_{1},\\& l_{10,6}=\frac{1}{2}u_{2}\partial^{-1}u_{5}+\frac{1}{2}u_{4}\partial^{-1}u_{6}+\frac{1}{2}u_{5}\partial^{-1}u_{2}+\frac{1}{2}u_{6}\partial^{-1}u_{4},\ l_{10,7}=-u_{2}\partial^{-1}u_{10}-u_{4}\partial^{-1}u_{7}-u_{6}\partial^{-1}u_{11},\\& l_{10,8}=-u_{1}\partial^{-1}u_{10}-u_{4}\partial^{-1}u_{8}-u_{5}\partial^{-1}u_{12},\ l_{10,10}=\frac{1}{2}\partial-\frac{1}{2}u_{1}\partial^{-1}u_{7}-\frac{1}{2}u_{2}\partial^{-1}u_{8}-u_{4}\partial^{-1}u_{10}-\frac{1}{2}u_{5}\partial^{-1}u_{11}-\frac{1}{2}u_{6}\partial^{-1}u_{12},\\& l_{10,11}=-\frac{1}{2}u_{2}\partial^{-1}u_{12}-\frac{1}{2}u_{4}\partial^{-1}u_{11}-\frac{1}{2}u_{6}\partial^{-1}u_{9},\ l_{10,12}=-\frac{1}{2}u_{1}\partial^{-1}u_{11}-\frac{1}{2}u_{4}\partial^{-1}u_{12}-\frac{1}{2}u_{5}\partial^{-1}u_{9},\\&
	l_{10,13}=-\frac{1}{2}u_{17}+\frac{1}{2}u_{2}\partial^{-1}u_{14}+\frac{1}{2}u_{4}\partial^{-1}u_{13}+\frac{1}{2}u_{6}\partial^{-1}u_{15},\ l_{10,14}=-\frac{1}{2}u_{16}+\frac{1}{2}u_{1}\partial^{-1}u_{13}+\frac{1}{2}u_{4}\partial^{-1}u_{14}+\frac{1}{2}u_{5}\partial^{-1}u_{15},\\& l_{10,16}=-\frac{1}{2}u_{1}\partial^{-1}u_{17}-\frac{1}{2}u_{4}\partial^{-1}u_{16},\ l_{10,17}=-\frac{1}{2}u_{2}\partial^{-1}u_{16}-\frac{1}{2}u_{4}\partial^{-1}u_{17},\ l_{10,18}=-\frac{1}{2}u_{5}\partial^{-1}u_{17}-\frac{1}{2}u_{6}\partial^{-1}u_{16},\\&
	l_{11,1}=u_{1}\partial^{-1}u_{5}+u_{5}\partial^{-1}u_{1},\ l_{11,2}=u_{4}\partial^{-1}u_{6}+u_{6}\partial^{-1}u_{4},\ l_{11,3}=u_{3}\partial^{-1}u_{5}+u_{5}\partial^{-1}u_{3},\\& l_{11,4}=\frac{1}{2}u_{1}\partial^{-1}u_{6}+\frac{1}{2}u_{4}\partial^{-1}u_{5}+\frac{1}{2}u_{5}\partial^{-1}u_{4}+\frac{1}{2}u_{6}\partial^{-1}u_{1},\ l_{11,5}=\frac{1}{2}u_{1}\partial^{-1}u_{3}+\frac{1}{2}u_{3}\partial^{-1}u_{1}+u_{5}\partial^{-1}u_{5},\\& l_{11,6}=\frac{1}{2}u_{3}\partial^{-1}u_{4}+\frac{1}{2}u_{4}\partial^{-1}u_{3}+\frac{1}{2}u_{5}\partial^{-1}u_{6}+\frac{1}{2}u_{6}\partial^{-1}u_{5},\ l_{11,7}=-u_{3}\partial^{-1}u_{11}-u_{5}\partial^{-1}u_{7}-u_{6}\partial^{-1}u_{10},\\& l_{11,9}=-u_{1}\partial^{-1}u_{11}-u_{4}\partial^{-1}u_{12}-u_{5}\partial^{-1}u_{9},\ l_{11,10}=-\frac{1}{2}u_{3}\partial^{-1}u_{12}-\frac{1}{2}u_{5}\partial^{-1}u_{10}-\frac{1}{2}u_{6}\partial^{-1}u_{8},\\& l_{11,11}=\frac{1}{2}\partial-\frac{1}{2}u_{1}\partial^{-1}u_{7}-\frac{1}{2}u_{3}\partial^{-1}u_{9}-\frac{1}{2}u_{4}\partial^{-1}u_{10}-u_{5}\partial^{-1}u_{11}-\frac{1}{2}u_{6}\partial^{-1}u_{12},\\& l_{11,12}=-\frac{1}{2}u_{1}\partial^{-1}u_{10}-\frac{1}{2}u_{4}\partial^{-1}u_{8}-\frac{1}{2}u_{5}\partial^{-1}u_{12},\\& l_{11,13}=-\frac{1}{2}u_{18}+\frac{1}{2}u_{3}\partial^{-1}u_{15}+\frac{1}{2}u_{5}\partial^{-1}u_{13}+\frac{1}{2}u_{6}\partial^{-1}u_{14},\ l_{11,15}=-\frac{1}{2}u_{16}+\frac{1}{2}u_{1}\partial^{-1}u_{13}+\frac{1}{2}u_{4}\partial^{-1}u_{14}+\frac{1}{2}u_{5}\partial^{-1}u_{15},\\& l_{11,16}=-\frac{1}{2}u_{1}\partial^{-1}u_{18}-\frac{1}{2}u_{5}\partial^{-1}u_{16},\ l_{11,17}=-\frac{1}{2}u_{4}\partial^{-1}u_{18}-\frac{1}{2}u_{6}\partial^{-1}u_{16},\ l_{11,18}=-\frac{1}{2}u_{3}\partial^{-1}u_{16}-\frac{1}{2}u_{5}\partial^{-1}u_{18},\\&
	l_{12,1}=u_{4}\partial^{-1}u_{5}+u_{5}\partial^{-1}u_{4},\ l_{12,2}=u_{2}\partial^{-1}u_{6}+u_{6}\partial^{-1}u_{2},\ l_{12,3}=u_{3}\partial^{-1}u_{6}+u_{6}\partial^{-1}u_{3},\\& l_{12,4}=\frac{1}{2}u_{2}\partial^{-1}u_{5}+\frac{1}{2}u_{4}\partial^{-1}u_{6}+\frac{1}{2}u_{5}\partial^{-1}u_{2}+\frac{1}{2}u_{6}\partial^{-1}u_{4},\ l_{12,5}=\frac{1}{2}u_{3}\partial^{-1}u_{4}+\frac{1}{2}u_{4}\partial^{-1}u_{3}+\frac{1}{2}u_{5}\partial^{-1}u_{6}+\frac{1}{2}u_{6}\partial^{-1}u_{5},\\& l_{12,6}=\frac{1}{2}u_{2}\partial^{-1}u_{3}+\frac{1}{2}u_{3}\partial^{-1}u_{2}+u_{6}\partial^{-1}u_{6},\ l_{12,8}=-u_{3}\partial^{-1}u_{12}-u_{5}\partial^{-1}u_{10}-u_{6}\partial^{-1}u_{8},\\& l_{12,9}=-u_{2}\partial^{-1}u_{12}-u_{4}\partial^{-1}u_{11}-u_{6}\partial^{-1}u_{9},\ l_{12,10}=-\frac{1}{2}u_{3}\partial^{-1}u_{11}-\frac{1}{2}u_{5}\partial^{-1}u_{7}-\frac{1}{2}u_{6}\partial^{-1}u_{10},\\& l_{12,11}=-\frac{1}{2}u_{2}\partial^{-1}u_{10}-\frac{1}{2}u_{4}\partial^{-1}u_{7}-\frac{1}{2}u_{6}\partial^{-1}u_{11},\\& l_{12,12}=\frac{1}{2}\partial-\frac{1}{2}u_{2}\partial^{-1}u_{8}-\frac{1}{2}u_{3}\partial^{-1}u_{9}-\frac{1}{2}u_{4}\partial^{-1}u_{10}-\frac{1}{2}u_{5}\partial^{-1}u_{11}-u_{6}\partial^{-1}u_{12},\\& l_{12,14}=-\frac{1}{2}u_{18}+\frac{1}{2}u_{3}\partial^{-1}u_{15}+\frac{1}{2}u_{5}\partial^{-1}u_{13}+\frac{1}{2}u_{6}\partial^{-1}u_{14},\ l_{12,15}=-\frac{1}{2}u_{17}+\frac{1}{2}u_{2}\partial^{-1}u_{14}+\frac{1}{2}u_{4}\partial^{-1}u_{13}+\frac{1}{2}u_{6}\partial^{-1}u_{15},\\& l_{12,16}=-\frac{1}{2}u_{4}\partial^{-1}u_{18}-\frac{1}{2}u_{5}\partial^{-1}u_{17},\ l_{12,17}=-\frac{1}{2}u_{2}\partial^{-1}u_{18}-\frac{1}{2}u_{6}\partial^{-1}u_{17},\ l_{12,18}=-\frac{1}{2}u_{3}\partial^{-1}u_{17}-\frac{1}{2}u_{6}\partial^{-1}u_{18},\\&
	l_{13,1}=-2u_{16}\partial^{-1}u_{1},\ l_{13,2}=-2u_{17}\partial^{-1}u_{4},\ l_{13,3}=-2u_{18}\partial^{-1}u_{5},\ l_{13,4}=-u_{16}\partial^{-1}u_{4}-u_{17}\partial^{-1}u_{1},\\& l_{13,5}=-u_{26}\partial^{-1}u_{5}-u_{18}\partial^{-1}u_{1},\ l_{13,6}=-u_{17}\partial^{-1}u_{5}-u_{18}\partial^{-1}u_{4},\ l_{13,7}=2u_{13}+2u_{16}\partial^{-1}u_{7}+2u_{17}\partial^{-1}u_{10}+2u_{18}\partial^{-1}u_{11},\\& l_{13,10}=u_{14}+u_{16}\partial^{-1}u_{10}+u_{17}\partial^{-1}u_{8}+u_{18}\partial^{-1}u_{12},\ l_{13,11}=u_{15}+u_{16}\partial^{-1}u_{11}+u_{17}\partial^{-1}u_{12}+u_{18}\partial^{-1}u_{9},\\& l_{13,13}=\partial-u_{16}\partial^{-1}u_{13}-u_{17}\partial^{-1}u_{14}-u_{18}\partial^{-1}u_{15},\ l_{13,16}=-u_{1}+2u_{16}\partial^{-1}u_{16},\ l_{13,17}=-u_{4}+u_{17}\partial^{-1}u_{16},\\& l_{13,18}=-u_{5}+u_{18}\partial^{-1}u_{16},\\&
	l_{14,1}=-2u_{16}\partial^{-1}u_{4},\ l_{14,2}=-2u_{17}\partial^{-1}u_{2},\ l_{14,3}=-2u_{18}\partial^{-1}u_{6},\ l_{14,4}=-u_{16}\partial^{-1}u_{2}-u_{17}\partial^{-1}u_{4},\\& l_{14,5}=-u_{16}\partial^{-1}u_{6}-u_{18}\partial^{-1}u_{4},\ l_{14,6}=-u_{17}\partial^{-1}u_{6}-u_{18}\partial^{-1}u_{2},\ l_{14,8}=2u_{14}+2u_{16}\partial^{-1}u_{10}+2u_{17}\partial^{-1}u_{8}+2u_{18}\partial^{-1}u_{12},\\& l_{14,10}=u_{13}+u_{16}\partial^{-1}u_{7}+u_{17}\partial^{-1}u_{10}+u_{18}\partial^{-1}u_{11},\ l_{14,12}=u_{15}+u_{16}\partial^{-1}u_{11}+u_{17}\partial^{-1}u_{12}+u_{18}\partial^{-1}u_{9},\\& l_{14,14}=\partial-u_{16}\partial^{-1}u_{13}-u_{17}\partial^{-1}u_{14}-u_{18}\partial^{-1}u_{15},\ l_{14,16}=-u_{4}+u_{16}\partial^{-1}u_{17},\ l_{14,17}=-u_{2}+u_{17}\partial^{-1}u_{17},\\& l_{14,18}=-u_{6}+u_{18}\partial^{-1}u_{17},\\&
	l_{15,1}=-2u_{16}\partial^{-1}u_{5},\ l_{15,2}=-2u_{17}\partial^{-1}u_{6},\ l_{15,3}=-2u_{18}\partial^{-1}u_{3},\ l_{15,4}=-u_{16}\partial^{-1}u_{6}-u_{17}\partial^{-1}u_{5},\\& l_{15,5}=-u_{16}\partial^{-1}u_{3}-u_{18}\partial^{-1}u_{5},\ l_{15,6}=-u_{17}\partial^{-1}u_{3}-u_{18}\partial^{-1}u_{6},\ l_{15,9}=2u_{15}+2u_{16}\partial^{-1}u_{11}+2u_{17}\partial^{-1}u_{12}+2u_{18}\partial^{-1}u_{9},\\& l_{15,11}=u_{13}+u_{16}\partial^{-1}u_{7}+u_{17}\partial^{-1}u_{10}+u_{18}\partial^{-1}u_{11},\ l_{15,12}=u_{14}+u_{16}\partial^{-1}u_{10}+u_{17}\partial^{-1}u_{8}+u_{18}\partial^{-1}u_{12},\\& l_{15,15}=\partial-u_{16}\partial^{-1}u_{13}-u_{17}\partial^{-1}u_{14}-u_{18}\partial^{-1}u_{15},\ l_{15,16}=-u_{5}+u_{16}\partial^{-1}u_{18},\ l_{15,17}=-u_{6}+u_{17}\partial^{-1}u_{18},\\& l_{15,18}=-u_{3}+u_{18}\partial^{-1}u_{18},\\&
	l_{16,1}=2u_{16}+2u_{13}\partial^{-1}u_{1}+2u_{14}\partial^{-1}u_{4}+2u_{15}\partial^{-1}u_{5},\ l_{16,4}=u_{17}+u_{13}\partial^{-1}u_{4}+u_{14}\partial^{-1}u_{2}+u_{15}\partial^{-1}u_{6},\\& l_{16,5}=u_{18}+u_{13}\partial^{-1}u_{5}+u_{14}\partial^{-1}u_{6}+u_{15}\partial^{-1}u_{3},\ l_{16,7}=-2u_{13}\partial^{-1}u_{7},\ l_{16,8}=-2u_{4}\partial^{-1}u_{10},\ l_{16,9}=-2u_{15}\partial^{-1}u_{11},\\& l_{16,10}=-u_{13}\partial^{-1}u_{10}-u_{14}\partial^{-1}u_{7},\ l_{16,11}=-u_{13}\partial^{-1}u_{11}-u_{15}\partial^{-1}u_{7},\ l_{16,12}=-u_{14}\partial^{-1}u_{11}-u_{15}\partial^{-1}u_{10},\\& l_{16,13}=u_{7}+u_{13}\partial^{-1}u_{13},\ l_{16,14}=u_{10}+u_{14}\partial^{-1}u_{13},\ l_{16,15}=u_{11}+u_{15}\partial^{-1}u_{13},\\& l_{16,16}=-\partial-u_{13}\partial^{-1}u_{16}-u_{14}\partial^{-1}u_{17}-u_{15}\partial^{-1}u_{18},\\&
	l_{17,2}=2u_{17}+2u_{13}\partial^{-1}u_{4}+2u_{14}\partial^{-1}u_{2}+2u_{15}\partial^{-1}u_{6},\ l_{17,4}=u_{16}+u_{13}\partial^{-1}u_{1}+u_{14}\partial^{-1}u_{4}+u_{15}\partial^{-1}u_{5},\\& l_{17,6}=u_{18}+u_{13}\partial^{-1}u_{5}+u_{14}\partial^{-1}u_{6}+u_{15}\partial^{-1}u_{3},\ l_{17,7}=-2u_{13}\partial^{-1}u_{10},\ l_{17,8}=-2u_{14}\partial^{-1}u_{8},\ l_{17,9}=-2u_{15}\partial^{-1}u_{12},\\& l_{17,10}=-u_{13}\partial^{-1}u_{8}-u_{14}\partial^{-1}u_{10},\ l_{17,11}=-u_{13}\partial^{-1}u_{12}-u_{15}\partial^{-1}u_{10},\ l_{17,12}=-u_{14}\partial^{-1}u_{12}-u_{15}\partial^{-1}u_{8},\\& l_{17,13}=u_{10}+u_{13}\partial^{-1}u_{14},\ l_{17,14}=u_{8}+u_{14}\partial^{-1}u_{14},\ l_{17,15}=u_{12}+u_{15}\partial^{-1}u_{14},\\& l_{17,17}=-\partial-u_{13}\partial^{-1}u_{16}-u_{14}\partial^{-1}u_{17}-u_{15}\partial^{-1}u_{18},\\&
	l_{18,3}=2u_{18}+2u_{13}\partial^{-1}u_{5}+2u_{14}\partial^{-1}u_{6}+2u_{15}\partial^{-1}u_{3},\ l_{18,5}=u_{16}+u_{13}\partial^{-1}u_{1}+u_{14}\partial^{-1}u_{4}+u_{15}\partial^{-1}u_{5},\\& l_{18,6}=u_{17}+u_{13}\partial^{-1}u_{4}+u_{14}\partial^{-1}u_{2}+u_{15}\partial^{-1}u_{6},\ l_{18,7}=-2u_{13}\partial^{-1}u_{11},\ l_{18,8}=-2u_{14}\partial^{-1}u_{12},\ l_{18,9}=-2u_{15}\partial^{-1}u_{9},\\& l_{18,10}=-u_{13}\partial^{-1}u_{12}-u_{14}\partial^{-1}u_{11},\ l_{18,11}=-u_{13}\partial^{-1}u_{9}-u_{15}\partial^{-1}u_{11},\ l_{18,12}=-u_{14}\partial^{-1}u_{9}-u_{15}\partial^{-1}u_{12},\\& l_{18,13}=u_{11}+u_{13}\partial^{-1}u_{15},\ l_{18,14}=u_{12}+u_{14}\partial^{-1}u_{15},\ l_{18,15}=u_{9}+u_{15}\partial^{-1}u_{15},\\& l_{18,18}=-\partial-u_{13}\partial^{-1}u_{16}-u_{14}\partial^{-1}u_{17}-u_{15}\partial^{-1}u_{18}.
	\end{align*}

\section*{bibliography}

\end{document}